\documentclass[acmsmall]{acmart}
\usepackage{macro}
\usepackage{graphicx}
\usepackage{subcaption}
\usepackage{graphicx}
\usepackage[most]{tcolorbox}

\newcommand{\name}{live}

\newcounter{packednmbr}

\newcounter{insightlabel}
\newcounter{insightnmbr}
\renewcommand{\theinsightlabel}{\textbf{\theinsightnmbr}}

\newcounter{challengelabel}
\newcounter{challengenmbr}
\renewcommand{\thechallengelabel}{\textbf{\thechallengenmbr}}

\newcounter{findinglabel}
\newcounter{findingnmbr}
\renewcommand{\thefindinglabel}{\textbf{\thefindingnmbr}}
\newenvironment{finding}{
\begin{list}{\textbf{Finding }\thefindinglabel:~}{\usecounter{findinglabel}\stepcounter{findingnmbr}\setlength{\labelwidth}{0pt}\setlength{\labelsep}{0pt}\setlength{\leftmargin}{0in}\noindent\rule{\linewidth}{1pt}\vspace{-2pt}\item \bf \em}}{\\[-7pt]\end{list}\vspace{-10pt}\noindent\rule{\linewidth}{1pt}}

\newcounter{limitationlabel}
\newcounter{limitationnmbr}
\renewcommand{\thelimitationlabel}{\textbf{\thelimitationnmbr}}

\newcommand{\commentout}[1]{}

\tikzset{
    circ/.style={circle, draw=black, text=black, fill=white,  minimum size=8mm},
    adv/.style={circle, draw=advc, text=black, fill=yellow!20!white,  minimum size=8mm},
    every picture/.style=very thick
}

\setcopyright{acmlicensed} 
\acmJournal{POMACS}
\acmYear{2023} 
\acmVolume{7} 
\acmNumber{2}
\acmArticle{32} 
\acmMonth{6} 
\acmPrice{15.00}
\acmDOI{10.1145/3589976}

\received{August 2022}
\received[revised]{February 2023}
\received[accepted]{April 2023}

\title{Strategic Latency Reduction in Blockchain Peer-to-Peer Networks}
\author{Weizhao Tang}
\affiliation{\institution{Carnegie Mellon Univerisity \& IC3}\country{USA}}
\email{wtang2@andrew.cmu.edu}
\author{Lucianna Kiffer}
\affiliation{\institution{ETH Zurich}\country{Switzerland}}
\email{lkiffer@ethz.ch}
\author{Giulia Fanti}
\affiliation{\institution{Carnegie Mellon Univerisity \& IC3}\country{USA}}
\email{gfanti@andrew.cmu.edu}
\author{Ari Juels$^*$}
\affiliation{\institution{Cornell Tech, IC3, \& Chainlink Labs\titlenote{The author contributed to this work in his Chainlink Labs role.}}\country{USA}}
\email{juels@cornell.edu}

\renewcommand{\shortauthors}{Weizhao Tang, Lucianna Kiffer, Giulia Fanti, \& Ari Juels}

\date{}

\begin{document}

\begin{abstract}
Most permissionless blockchain networks run on peer-to-peer (P2P) networks, which offer flexibility and decentralization at the expense of performance (e.g., network latency). Historically, this tradeoff has not been a bottleneck for most blockchains. However, an emerging host of blockchain-based applications (e.g., decentralized finance) are increasingly sensitive to latency; users who can reduce their network latency relative to other users can accrue (sometimes significant) financial gains. 

In this work, we initiate the study of strategic latency reduction in blockchain P2P networks. We first define two classes of latency that are of interest in blockchain applications. We then show empirically that a strategic agent who controls only their local peering decisions can manipulate both types of latency, achieving 60\% of the global latency gains provided by the centralized, paid service bloXroute, or, in targeted scenarios, comparable gains. Finally, we show that our results are not due to the poor design of existing P2P networks. Under a simple network model, we theoretically prove that an adversary can always manipulate the P2P network's latency to their advantage, provided the network experiences sufficient peer churn and transaction activity. 
\end{abstract}




\maketitle


\section{Introduction}
\label{sec:intro}







Most permissionless blockchains today run on peer-to-peer (P2P) communication networks due to their flexibility and distributed nature \cite{li2005routing,delgado2018cryptocurrency}.
These benefits of P2P networks typically come at the expense of network performance, particularly the latency of message delivery \cite{p2p,delgado2018cryptocurrency,decker2013information}.
Historically, this has not been a bottleneck because most permissionless blockchains are  currently performance-limited not by network latency, but by the rate of block production and transaction confirmation, both of which are \resub{superlinear in network latency and} dictated by the underlying consensus mechanism \cite{garay2015bitcoin,prism,yu2020ohie,li2020decentralized,decker2013information}.
For this reason, network latency has not traditionally been a first-order concern in blockchain P2P networks, except where substantial latency on the order of many seconds or even minutes raises the risk of forking~\cite{croman2016scaling}.

Emerging concerns in blockchain systems, however, are beginning to highlight the importance of \textit{variations in fine-grained network latency}---e.g., on the order of milliseconds---among nodes in blockchain P2P networks. These concerns are largely independent of the underlying consensus mechanism. They revolve instead around strategic behaviors that arise particularly in smart-contract-enabled blockchains, such as Ethereum, where decentralized finance (DeFi) applications create timing-based opportunities for financial gain.
Key examples of such opportunities include:
\begin{itemize}[leftmargin=*]
    \item \textbf{Arbitrage:} Many blockchain systems offer highly profitable opportunities for \textit{arbitrage}, in which assets are strategically sold and bought at different prices on different markets (or at different times) to take advantage of price differences~\cite{makarov2020trading}.
          Such arbitrage can involve trading one cryptocurrency for another, buying / selling mispriced non-fungible tokens (NFTs), etc., on blockchains and/or in  centralized cryptocurrency exchanges.  Strategic agents performing arbitrage can obtain an important advantage through low latency access to blockchain transactions.
    \item \textbf{Strategic transaction ordering:} In many public blockchains, strategic agents \sigm{(who do not validate blocks)} can profit by ordering their transactions in ways that exploit other users---a phenomenon referred to as \textit{Miner-Extractable Value} (MEV)~\cite{daian2020flash,FlashbotsDashboard}. For example, strategic agents may  \textit{front-run} other users' transactions, i.e., execute strategic transactions immediately before those of victims, or \textit{back-run} key transactions---e.g., oracle report delivery or token sales---to gain priority in transaction ordering.
    Small network latency reductions can allow an adversary to observe victims' transactions before competing strategic agents do. 
    \cam{
        With more adversarial nodes and links available, it may even attempt to choose peers in order to frontrun as many pairs of sources and destinations as possible. 
    }
    \item \textbf{Improved block composition:} Miners (or validators) rely  on P2P networks to observe user transactions, which they then include in the blocks they produce. Which transactions a miner includes in a block determines the fees it receives and thus its profits. Therefore the lower the latency a miner experiences in obtaining new transactions, the higher its potential profit.\footnote{Receiving blocks from other miners quickly is also beneficial. To reduce forking risks, however, miners or validators often bypass P2P networks and instead use cut-through networks to communicate blocks to one another~\cite{FIBRE:2022}.}
    \item \textbf{Targeted attacks:} The security of nodes in blockchain P2P networks may be weakened or compromised by adversaries' discovery of their IP addresses. For example, a node may issue critical transactions that update asset prices in a DeFi smart contract. Discovery of its IP address could result in  denial-of-service (DoS) attacks against it. Similarly, a user who uses her own node to issue transactions that suggest possession of a large amount of cryptocurrency could be victimized by targeted cyberattacks should her IP address be revealed.
\end{itemize}
  We show that small differences in transaction latencies can help an adversary distinguish between paths to a victim---and ultimately even discover the victim's IP address.
There is a strong incentive, therefore, for agents to minimize the latency they experience in P2P networks.
\textbf{In this work, we initiate the study of strategic latency reduction in blockchain P2P networks}. 

\subsection{Types of network latency}

In our investigations, we explore two types of latency that play a key role in blockchain P2P networks: \textit{direct latency} and \textit{triangular latency} (formal definitions in \sref{sec:model}).
\sigm{We demonstrate these types of latencies in Table \ref{tab:landscape}. }

\bigskip
\noindent \textbf{Direct latency}
refers to the latency with which messages reach a listener node from one or more vantage points---in other words, source latency. We consider two variants:
\begin{itemize}[leftmargin=*]
    \item \emph{Direct targeted latency} is the delay between a transaction being pushed onto the network by a specific target node (the \emph{victim}) and a node belonging to a given agent receiving the transaction, averaged across all transactions produced by the victim. We can view targeted latency either in terms of an absolute latency, or a relative latency compared to other nodes receiving the same transaction; we focus in this paper on relative latency.
    \item \emph{Direct global latency} for a given agent's node refers to targeted latency for that node averaged  over all source nodes in the network. In other words, there is no single target victim.
\end{itemize}

\smallskip
\noindent \textbf{Triangular latency} is a second type of latency that corresponds to a node's ability to inject itself between a pair of communicating nodes. Low triangular latency is motivated by (for example) a node's desire to front-run another node's transactions.
We consider two forms of triangular latency:
\begin{itemize}[leftmargin=*]
    \item \emph{Triangular targeted latency} refers to the ability of an agent's node $v$ to ``shortcut" paths between a sender $s$ (e.g., the creator of a transaction) and a receiver $r$ (e.g., a miner or miners). That is, suppose $s$ creates a transaction $m$ that is meant to reach a (set of) target nodes $r$. Triangular targeted latency measures the difference between the total delay on path $s \to v \to r$ and the smallest delay over all paths from $s \to r$ not involving $v$. Given negative triangular targeted latency, an agent can front-run successfully by injecting a competing transaction $m'$ into $v$ upon seeing a victim's transaction $m$.
    \item \emph{Triangular global latency} refers to the triangular relative targeted latency averaged over all source-destination pairs in the network.
\end{itemize}

Direct latency and triangular latency are related, but not identical, as we explain in \sref{sec:model} and \sref{sec:triangular-latency}.

\subsection{Latency-reduction strategies}

High-frequency trading (HFT) firms in the traditional finance industry compete aggressively to reduce the latency of their trading platforms and connections to markets---in some cases, by nanoseconds---using technologies like hollow-core fiber optics~\cite{Osipovich:2020} and satellite links~\cite{Osipovich:2021}.

Such approaches are not viable for blockchain P2P networks, in which \emph{network topology} matters more than individual nodes' hardware configurations. These networks are also permissionless and experience high churn rates, and are controlled by different parties. 

In practice, nodes may rely on proprietary, \textit{cut-through} networks to learn transactions or blocks quickly. Miners maintain private networks~\cite{FIBRE:2022}. There are also public, paid cut-through networks, such as bloXroute~\cite{BloXroute}.
An agent in a blockchain P2P network, however, can also act strategically by means of \textit{local actions, namely choosing which peers a node under its control connects to}.

\smallskip
\noindent{\textbf{Peri:}} A recent protocol called Perigee~\cite{mao2020perigee} leverages agents' ability in blockchain P2P networks to control their peering to achieve network-wide latency improvements. In the Perigee protocol, \textit{every} node favors peers that relay blocks quickly and rejects peers that fail to do so.
We observe that Perigee-like strategies can instead be adopted by \textit{individual strategic agents}.
We adopt a set of latency-reduction strategies called \textbf{\textit{Peri}}\footnote{A mischievous winged spirit in Persian lore.} that modify Perigee with optimizations tailored for the individual-agent setting.
In particular, we show that agents using Peri can gain a latency advantage over other agents in both direct and triangular latency, including targeted and global variants. 

\subsection{Our contributions}
We explore techniques for an agent to reduce direct measurement latency and triangular latency in a blockchain P2P network using only \emph{local peering choices}, i.e., those of node(s) controlled by the agent.
\textbf{Overall, we find that strategic latency reduction is possible, both in theory and in practice.}
Our contributions are as follows. 
\begin{itemize}[leftmargin=*]

    \item \textbf{Practical strategic peering:} We demonstrate empirically and in simulation that the strategic peering scheme Peri can achieve direct latency improvement compared to the current Ethereum P2P network.\footnote{We explicitly do not consider the design of \peri to be a main contribution, as the design is effectively the same as Perigee. We make a distinction between the two purely to highlight their differing goals and implementation details.} We instrument the \texttt{geth} Ethereum client to measure direct measurement latency and implement Peri. We observe direct global latency and direct targeted latency reductions each of about 11 ms---over half the reduction observed for bloXroute for direct latency and comparable to that of bloXroute's paid service for targeted latency. 
    We additionally show that Peri can discover the IP address of a victim node 7$\times$ more frequently than baseline approaches. 
    \item \textbf{Hardness of triangular-latency minimization:} We explore the question of strategic triangular-latency reduction via a graph-theoretic model. We show that solving this problem optimally is NP-hard, and the greedy algorithm cannot approximate the global solution. 
    These graph-theoretic results may be of independent interest.
    We show experimentally in simulations, however, that Peri \textit{is} effective in reducing such latency.

    \item \textbf{Impossiblity of \fairness:} 
    Within a theoretical model, we prove that  \textit{strategy-proof P2P network design is fundamentally unachievable}: For any default peering algorithm used by nodes in a P2P network, as long as the network experiences
    natural churn and a target node is active, a strategic agent can always reduce direct targeted latency relative to agents following the default peering algorithm. 
    Specifically, with probability at least $1 - \epsilon$ for any $\epsilon \in (0, 1)$,
    an agent can connect directly to a victim in time $O(\epsilon^{-1} \log^2 (\epsilon^{-1}))$. 
%

\end{itemize}

\section{Background and Related Work}
\label{sec:background}

Many of the latency-sensitive applications discussed in \sref{sec:intro} arise in decentralized finance (DeFi) and require a blockchain platform that supports general smart contracts. 
We therefore focus on the Ethereum network \sigm{(in particular, the {eth1} or the execution layer)}; however, 
the concepts and analysis apply to other blockchain P2P networks. 

\paragraph{Network Formation}
Ethereum, like most permissionless blockchains, maintains a P2P network among its nodes.
Each node is represented by its enode ID, which encodes the node's IP address and TCP and UDP ports \cite{wang2021ethna}. Nodes in the network learn about each other via a node discovery protocol based on the Kademlia distributed hash table (DHT) protocol~\cite{maymounkov2002kademlia}. To bootstrap after a quiescent period or upon first joining the network, a node either queries its previous peers or hard-coded bootstrap peers about other nodes in the network. Specifically, it sends a \texttt{FINDNODE} request using its own enode ID as the DHT query seed. The node's peers respond with the enode IDs and IP addresses of those nodes in their own peer tables that have IDs closest in distance to the query ID.
Nodes use these responses to populate their local peer tables and identify new peers. 

\paragraph{Transaction Propagation}
When a user wants to make a transaction, they first cryptographically sign the transaction using a fixed public key, then broadcast the signed transaction over the P2P network using a simple flooding protocol~\cite{ethwire}.
Each node $v$, upon seeing a new transaction $m$, first checks the transaction's validity; if the transaction passes validity checks, it is added to $v$'s local \texttt{TxPool} of unconfirmed transactions.
Then, $v$ executes a simple three-step process: (1) It chooses a small random subset $\{u_i\}$ of its connected peer nodes; (2) $v$ sends a transaction hash $H(m)$ to each peer $u_i$; (3) If $u_i$ has already seen transaction $m$, it communicates this to the sender $v$; otherwise it responds with a \texttt{GetTx} request, and $v$ in turn sends $m$ in full \cite{wang2021ethna}.
There are two main sources of latency in the P2P network: (1) network latency, which stems from sending messages over a P2P overlay of the public Internet, (2) node latency, which stems from local computation at each node before relaying a transaction. 
We treat node latency as fixed and try to manipulate network latency. 

\paragraph{Transaction Confirmation}
After transactions are disseminated over the P2P network, \emph{miner} nodes confirm transactions from the \texttt{TxPool} by compiling them into blocks \cite{monrat2019survey,wang2021ethna}. 
When forming blocks, miners choose the order in which transactions will be executed in the final ledger. 
This ordering can have significant financial repercussions (\sref{sec:applications}). 
Miners typically choose the ordering of transactions in a block based on a combination of (a) transactions' time of arrival, and (b) incentives and fees associated with mining a transaction in a particular order. 
In Ethereum, each block has a base fee, which is the minimum cost for being included in the block. In addition, transaction creators add a tip (priority fee), which is paid to the miner to incentivize inclusion of the transaction in a block \cite{fees,roughgarden2020transaction}. 
These fees are commonly called \emph{gas fees} in the Ethereum ecosystem.



\subsubsection{Example: The role of latency in arbitrage}
\label{sec:applications}
Among the applications in \sref{sec:intro}, the interplay between network latency and arbitrage is particularly delicate. 
To perform successful arbitrage, the arbitrageur must often make a front-running transaction $f$ immediately before the target transaction $m$, i.e., the transactions $f$ and $m$ are mined within the same block, but $f$ is ordered before $m$.
Techniques for achieving this goal have shifted over time. 


Before 2020, arbitrage on the Ethereum blockchain happened mostly via priority gas auctions (PGAs), in which arbitrageurs would observe a victim transaction, then publicly broadcast front-running transactions with progressively higher gas prices \cite{daian2020flash}. 
The arbitrageur with the highest gas price would win.
It benefits an arbitrageur to have a low triangular targeted latency (Definition \ref{def:latrt}) between target source nodes (e.g., the victim, other arbitrageurs) and a destination miner, or a low direct targeted latency (Definition \ref{def:latt}) to a miner. 
This allows swifter reactions to competing bids and more rebidding opportunities before the block is finalized.

In 2020, mechanisms for arbitrage began to shift to private auction channels like Flashbots \cite{FlashbotsDashboard}, in which an arbitrageur submits a miner-extractable-value (MEV) bundle with a tip for miners to a public relay, which privately forwards the bundle to miners.
A typical bundle consists of a front-running transaction and the target transaction, where the order of transactions is decided by the creator and cannot be changed by the miner of the bundle.
Among competing bundles (which conflict by including the same victim transaction), the one with the highest tip is mined.

The key difference from PGA is that an arbitrageur is no longer aware of the tips of its competitors;
tips are chosen blindly to balance cost and chance of success.
However, the tip is upper-bounded for a rational arbitrageur, because the arbitraging gain is determined by the victim transaction itself.
Competing arbitrageurs may set up a grim trigger 
\sigm{\footnote{\sigm{Grim trigger is a trigger strategy in which a player begins by cooperating in the first period, and continues to cooperate until a single defection by her opponent, after which the player defects forever.}}} 
on the percentage of the tip compared to the arbitrage profit \cite{daian2020flash}.
For instance, they may agree that the tip must not exceed 80\% of the profit, so that 20\% of profit is guaranteed for the winner.
Assuming that every arbitrageur is paying the miner with the same maximum tip, the competition collapses to one of speed.
Therefore, even with MEV bundles, 
network latency remains a critical component of arbitrage success.




\subsection{Related Work}
\label{sec:related}

Reducing P2P network latency has been a topic of significant interest in the blockchain community, including  
(1) decentralized protocol changes, and (2) centralized relay networks. 
 
\paragraph{Decentralized Protocol Changes}
Various decentralized protocols have been proposed to reduce the latency of propagating blocks and/or transactions. 
Several of these focus on reducing bandwidth usage, and reap latency benefits as a secondary effect. 
For example, Erlay \cite{naumenko2019erlay} uses set reconciliation to reduce bandwidth costs of transaction propagation, and Shrec \cite{han2020shrec} further designs a new encoding of transactions and a new relay protocol. 
These approaches are complementary to our results, which are not unique to the broadcast protocol or \sigm{compressive} transaction encoding scheme. 

Another body of work has attempted to bypass the effects of latency and peer churn on the topology of the P2P network by creating highly-structured networks and/or propagation paths while maintaining an open network; examples include Overchain and KadCast \cite{aradhya2022overchain,toshniwal2021comparative,rohrer2019kadcast}. 
Such highly-structured networks would be resilient to simple peering-choice strategies like \peri.

\paragraph{Centralized Relay Networks}
There have been multiple efforts to build centrally-operated, geographically-distributed relay networks for blockchain transactions and blocks. 
These include bloXroute \cite{BloXroute}, Fibre \cite{FIBRE:2022}, Bitcoin Relay Network \cite{falcon}, and the Falcon network \cite{falcon}.
In this work, we evaluate the latency reduction of centralized services, using bloXroute as a representative example that operates on the Ethereum blockchain. 
Nodes in the Ethereum P2P network connect to the bloXroute relay network via a gateway, which can either be local or hosted in the cloud \cite{bloxroute-docs}. 
BloXroute offers different tiers of service, which affect the resources available to a subscriber.

\section{Model}
\label{sec:model}


We \resub{begin by modelling} the P2P network $\Nc$ as a (possibly weighted) graph $(\Vc, \Ec)$.
\resub{Edge weights represent the physical distance traversed by a packet traveling between a pair of nodes (e.g., this can be approximated in practice with traceroute). In 
\sref{sec:triangular-latency}, we consider an unweighted, simplified model for analytical tractability.}
Let $d_{\Nc}(s, t)$ denote the shortest distance between $s$ and $t$ over the graph $\Nc$.
Each individual node $v \in \Vc$ can create a message $m$ and broadcast it to the network.
The message $m$ traverses all the nodes in $\Vc$ following the network protocol,
which allows an arbitrary node to forward the message $m$ upon receipt to a subset of its neighbors. 
We assume the source nodes of these messages follow a distribution $\Sb$ with support $\Vc$, i.e., $\Sb$ specifies a probability distribution over transaction emissions by nodes in $\Vc$.

A participant of the P2P network has two classes of identifiers:
\begin{enumerate}[label=Class \arabic*.]
    \item \textbf{(Network ID)} An ID assigned uniquely to each of its nodes. 
    For example, each node in the Ethereum P2P network is assigned a unique enode ID including the IP address.
    For simplicity, we let $\Vc$ denote the ID space, and a node $v \in \Vc$ represents an ID instance. 

    \item \textbf{(Logical ID)} 
    An ID that is used to identify the creator or owner of a message (e.g., the public key of a wallet). 
    We let $\Wc$ denote the ID space. 
\end{enumerate}

In blockchain networks, a participant may own multiple instances of both classes.
In our analysis, we assume there exists a mapping $\mathtt{NID}: \Wc \to \Vc$ from a logical ID to a network ID for simplicity.

\subsection{Adversarial Model}
\label{sec:adversarial-model}
\resub{
    We consider an adversary who aims to reduce latency in order to gain profit and/or threaten network security, as mentioned in \sref{sec:intro}. 
    As a starting point, we assume the adversary is capable of inserting one agent node $\agentnode$ into the network, which can maintain at most $k$ peer connections at a time.\footnote{\sigm{A motivated attacker can cheaply connect to many peers. However, in practice, we find empirically that maintaining too many peer connections is difficult due to issues including peer discovery and message processing, to name a few \cite{lighthouse-book}.}}
    In addition to receiving and sending messages like a normal node, the node $\agentnode$ can behave arbitrarily during relaying; for instance, it can block messages when the protocol requires it to broadcast them. 
    The adversary observes the network traffic through $\agentnode$ consisting of transactions, where each transaction has a signer, i.e., the logical ID of its sender. 
    The adversary is assumed to not know the mapping $\mathtt{NID}$, and cannot peer to a transaction sender $v = \mathtt{NID}(w)$ with only knowledge of the logical ID $w$. 
    With the network ID $v$, however, the adversary may establish a peer connection between $\agentnode$ and the node. 
}

\resub{
    We consider multiple adversarial models with the above capabilities, but different goals. 
    In particular, they aim to minimize different types of latencies, which we term direct and triangular latencies, with both global and targeted variants. 
    We start by precisely defining these latencies. 
}


In Table \ref{tab:landscape}, we show the relationship between these metrics, and summarize what is currently known (and unknown) about how to optimize them.
In this figure, solid red nodes represent a strategic agent.
Blue striped nodes represent target nodes in targeted latency metrics. 
Red thick lines represent the edges (peer connections) that can be formed by the agent to attempt to optimize its network latency. 


\begin{table}[]
    \centering \footnotesize \renewcommand{\arraystretch}{1.8}
    \begin{tabular}{|m{.15\linewidth}<{\centering}|*{2}{m{.38\linewidth}<{\centering}|}}
    \hline 
        \multicolumn{3}{|c|}{
            $\begin{array}{l}\includegraphics[height=1.2\mheight]{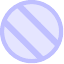} \end{array}$ \hspace{-.5em} Target node(s) ~~ 
            $\begin{array}{l}\includegraphics[height=1.2\mheight]{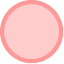} \end{array}$ \hspace{-.5em} Agent node(s) ~~ 
            $\begin{array}{l}\includegraphics[height=1.2\mheight]{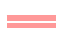} \end{array}$ \hspace{-.5em} Agent-initiated connections } 
            \\
    \hline 
        & \renewcommand{\arraystretch}{2} \textbf{\small Targeted} & \textbf{\small Global} \\
    \hline
        \textbf{\small Direct Latency} & \includegraphics[width=.7\linewidth]{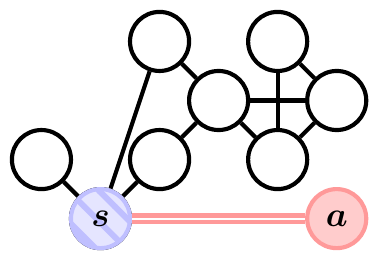} & \includegraphics[width=.7\linewidth]{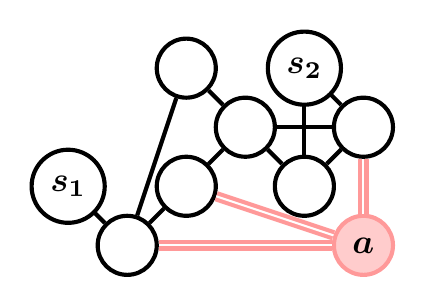} \\[-1ex]
    \hline
        \rowcolor{lightgray!30!white}
        \textbf{\small Optimized by} & Directly connecting to target  (\sref{sec:direct-metrics}) & SS-ASPDM approximation algorithm \cite{meyerson2009minimizing} \\
    \hline
        \textbf{\small Triangular Latency} & \includegraphics[width=.7\linewidth]{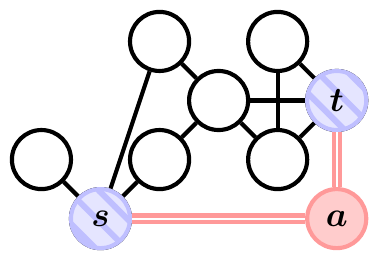} & \includegraphics[width=.7\linewidth]{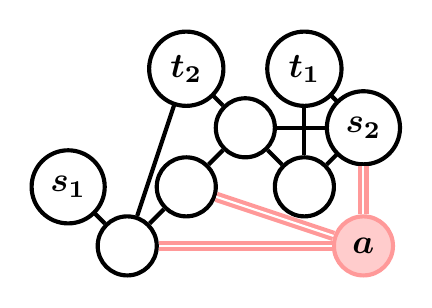} \\[-1ex]
    \hline
        \rowcolor{lightgray!30!white}
        \textbf{\small Optimized by} & Directly connecting to targets  (\sref{sec:triangular-metrics}) & Unknown (\sref{sec:triangular-hardness}) \\
    \hline
        \textbf{\small Main Challenge} & Don't know the IP addresses of targets (\sref{sec:unfair-infer}) & Don't know the graph topology, computational hardness of algorithms (\sref{sec:direct-metrics}, \sref{sec:triangular-hardness}) \\
    \hline
    \end{tabular}
    \caption{Challenges and algorithms for optimizing direct and triangular latency, both targeted and global. \vspace{-5ex}}
    \label{tab:landscape}
\end{table}

We let the random variable $\Lambda(\agentnode)$ denote the end-to-end traveling time of a message to $\agentnode$ from a random source $S \sim \Sb$; \resub{the randomness is over the source and (possibly) the network relay protocol. }
Further, we let $\Lambda_v(\agentnode)$ denote the random variable $\Lambda(\agentnode) | S = v$, representing end-to-end traveling time of a message from a single source $v$ to $\agentnode$. 
\resub{The randomness of $\Lambda_v(\agentnode)$ is over the network relay protocol.}
\resub{
    We measure the distributions of both random variables in \sref{sec:dml}. 
}


\subsection{Direct Latency}
\label{sec:direct-metrics}

Direct targeted latency (\dref{def:latt}) is defined as the expected single-source message travel time $\Lambda_v$.\footnote{\sigm{Expectation is only one metric of interest; practitioners may care about other metrics, e.g., quantiles. While we include such metrics in our experiments, we find that expectation captures the desired properties, while also facilitating basic analysis.}} 

\begin{definition}
    In a P2P network $\Nc$, the \textbf{\latt{}} of node $\agentnode$ with respect to a target node $v$ is defined as:
$$L_v(\agentnode) \triangleq \E[\Lambda_v(\agentnode)].$$
    \label{def:latt}
\end{definition}
Reducing \latt{} can help with applications such as targeted attacks or front-running a specific set of victim nodes.
We define \latg{} as follows.


\begin{definition}
    In a P2P network $\Nc$, the \textbf{\latg{}} of node $\agentnode$ is defined as: 
    \begin{equation*}
        L(\agentnode) \triangleq \E[\Lambda(\agentnode)].
    \end{equation*}
    \label{def:latg}
\end{definition}

Note that $L(\agentnode) = \E_{S \sim \Sb}[L_S(\agentnode)]$, which means the \latg{} equals the average \latt{} weighted by traffic source distribution $\Sb$. 


\subsubsection{Optimizing Direct Latency}

We can consider the problem of latency reduction by an agent as an optimization problem.
As a simplification to provide basic insights, we assume the network topology is fixed, and the end-to-end delay of an arbitrary message from source $s$ to destination $t$ is proportional to $d_{\resub{\Nc}}(s, t)$, which denotes the \resub{graph} distance between $s$ and $t$ on the network \resub{(i.e., number of hops, weighted by edge weights)}.
We also assume the agent has a budget of peer links $\peerbudget \ge 1$.
Then, we aim to solve the following problem to optimize \latt{}: 
\begin{align}
    \minimize \quad & L_v(\agentnode)                                      \notag              \\
    \sjt \quad      & \abs{ \brc{y \,|\, (\agentnode, y) \in \Ec} } \le \peerbudget. \label{eqn:dtl_opt}
\end{align}

\sigm{If we simplify the problem by assumption $d_{\resub{\Nc}}(s, t) \le d_{\resub{\Nc}}(s, r) + d_{\resub{\Nc}}(r, t)$ for all $r, s, t \in \Vc$, the solution to optimization \eqref{eqn:dtl_opt} becomes trivial}.
The agent only needs to add the target $v$ as a peer in order to achieve the minimum targeted latency, as stated in Table \ref{tab:landscape}. 
However, achieving this is not necessarily trivial:  an agent may not know $v$'s network address (e.g., IP address), even if $v$'s logical address is known. 
We discuss how to circumvent this challenge in \sref{sec:unfair-direct-targeted} \sigm{and what happens in practice when the simplifying assumption does not hold in \sref{sec:dml:targeted}}.

The situation is very different with the optimization problem for global latency reduction, which we formulate as follows.
\delresub{
Let $w_v$ denote the probability of $v$ being a target source, i.e., a source of messages of interest to the agent. 
}
The optimization problem is:
\begin{align*}
    \minimize \quad & L(\agentnode) = \sum_{v \in \Vc - \{\agentnode\}} 
    d(v, \agentnode) \cdot \resub{\P_{S \sim \Sb}[S = v]} \\
    \sjt \quad      & \abs{ \brc{y \,|\, (\agentnode, y) \in \Ec} } \le \peerbudget.
\end{align*}

This problem is called single-source average shortest path distance minimization (SS-ASPDM)~\cite{meyerson2009minimizing}.
It is NP-hard, but has an $\alpha$-approximate algorithm \cite{meyerson2009minimizing}.

\subsection{Triangular Latency}
\label{sec:triangular-metrics}

Triangular latency is motivated by applications related to front-running in P2P networks. 
It is defined with respect to one or more pairs of target nodes, where a pair of target nodes includes a source node $s$ and destination node $t$. 
For example, consider Table \ref{tab:landscape} (Triangular Targeted Latency), and let $s$ represent the creator of a transaction and $t$ a miner. 
The goal of agent node $\agentnode$ is to establish a path $s \to \agentnode \to t$ with lower travel time than any path on $\Nc$ from $s$ to $t$ excluding $\agentnode$. 
It can do so by adding edges to the P2P network, which are shown in red in Table \ref{tab:landscape}; note that the (shortest) path from $s$ to $a$ to $t$ is four hops, whereas the shortest path from $s$ to $t$ excluding $a$ is five hops.
The existence of such a path enables $\agentnode$ to front-run $s$'s transactions that are mined by $t$. 

We let $L_u'$ denote the \latt{} with respect to $u$ on the network excluding the agent $\agentnode$ and its incident edges.
As a front-runner, $\agentnode$ aims to satisfy the following condition:
\begin{equation}
    L'_s(t) > L_s(\agentnode) + L_t(\agentnode). \label{eqn:frineq}
\end{equation}
Here we assume the symmetry of targeted latency, i.e., $L_t(\agentnode) = L_a(t)$.
The left-hand side $L'_s(t)$ is a constant, while the right-hand side depends on the neighbors of $\agentnode$ over the P2P network $\Nc = (\Vc, \Ec)$; we define this quantity as \latrt{}.

\begin{definition}
    In a P2P network $\Nc$, the \textbf{\latrt{}} of node $\agentnode$ with respect to a target pair $(s,t)$ is defined as
    $$
        L_{s,t}(\agentnode) \triangleq L_s(\agentnode) + L_t(\agentnode).
    $$
    \label{def:latrt}
\end{definition}
\vspace{-4ex}
We again optimize this subject to a constraint on the number of agent peer connections: 
\begin{align}
    \minimize \quad & L_{s,t}(\agentnode)                                      \notag              \\
    \sjt \quad      & \abs{ \brc{y \,|\, (\agentnode, y) \in \Ec} } \le \peerbudget. \label{eqn:ttl_opt}
\end{align}
The optimal strategy for solving \eqref{eqn:ttl_opt} is again trivial: the agent should connect to both $s$ and $t$ (Table \ref{tab:landscape}). 
Hence, to optimize for a single source and destination, it is key to find $s$ and $t$ on the network and to ensure \eqref{eqn:frineq} holds. We discuss this further in \sref{sec:frl:impossibilty}.

We next consider an agent that tries to manipulate the \textit{global} triangular latency, capturing front-running opportunities over the entire network.
We start with a simple (and unsatisfactory) definition of \latrg{}.
Let $Q$ denote the set of source-destination pairs of network traffic.
A front-running agent 
might try to optimize the following:
\begin{equation}
L_Q(\agentnode) \triangleq \sum_{(s, t) \in Q} L_{s, t}(\agentnode) = \sum_{(s, t) \in Q} L_s(\agentnode) + L_t(\agentnode).
\label{eqn:naive_global_triangular}
\end{equation}

This is a variation of the SS-ASPDM problem discussed previously in \sref{sec:direct-metrics}.
Note, however, that an agent with minimal aggregated pairwise triangular latency $L_Q$ does \textit{not necessarily gain maximum profit from front-running}.
To profit, the agent needs to ensure the inequality~\eqref{eqn:frineq} holds for pairs $(s,t)$
not necessarily that right-hand side in~\eqref{eqn:naive_global_triangular} is minimized. 

We thus define the following proxy metric for \latrg{}, which instead measures the quality of peering choices made by front-runners.
Intuitively, the metric counts the number of node pairs $(s,t)$ that an adversarial agent can shortcut,
and the agent wants it to be as high as possible.
For example, in Table \ref{tab:landscape} (Triangular Global Latency), the agent $a$ is allowed to add three edges to the network, and its goal is to select those edges so as to shortcut the maximum number of node pairs $(s,t)$ from network $\Nc$.

\begin{definition}
    Let $\agentnode \notin \Vc$ denote an adversarial agent node, and $U \subseteq \Vc$ the set of $\agentnode$'s peers.
    %
    $\Nc' = (\Vc', \Ec')$ represents the network modified by the agent $a$, where $\Vc' = \Vc \cup \{ \agentnode \}$ and $\Ec' = \Ec \cup \bigcup_{u \in U} (u, \agentnode)$.
    Let $\Sc, \Tc \subseteq \Vc$ be the sets of sources and destinations, respectively.
    The agent node $\agentnode$ has a static distance penalty $\tau$, which means that it can only successfully front-run a source-destination pair $(s,t)$ if the path $s\to \agentnode \to t$ is at least $\tau$ units shorter than the shortest path $s\to t$ on $\Nc$ that does not pass through $\agentnode$.
    The \textbf{adversarial advantage} is defined as
    \begin{align}
        A_{\Nc}(U) = \sum_{s \in \Sc, t \in \Tc} 
        \bigg( & \I\big[d_{\Nc'}(s, t) + \tau < d_{\Nc}(s, t)\big] + 
                     \frac{1}{2} \cdot \I\big[d_{\Nc'}(s, t) + \tau = d_{\Nc}(s, t)\big] \bigg).
    \label{eqn:adversarial-advantage}
    \end{align}
    \label{def:advantage}
\end{definition}


\vskip-2ex
\resub{Note that we avoid defining the advantage with random end-to-end latencies such as $\Lambda_s(t)$ in order to keep the problem theoretically tractable. 
We also assign equal weights to each source-destination pair following the assumption that the adversary is unable to predict the profit to shortcut each particular pair of sources and destinations.  
}
In practice, $\Tc$ is the set of miners, and $\tau \in [0, \infty)$ is a parameter that depends on the computational capabilities of the agent, as well as randomness in the network.
Ideally, we assume $\Nc$ is uniformly weighted.
We also assume the weights of adversarial links $d_{\Nc'}(u, a) = 0$ for all $u \in U$, so that we have maximum flexibility in controlling the time costs over these links with parameter $\tau$. 

\paragraph{Summary}
The proposed latency metrics  express natural strategic agent goals. 
They are difficult to optimize directly, however, due to a combination of strong knowledge requirements about the P2P network, high computational complexity, and/or assumptions about the stationarity of the underlying network.
For example, to optimize global latency, we require two key assumptions that are unrealistic for P2P networks: the network topology should be static and the agent should know both the network topology and traffic patterns.
To optimize targeted latency, we require knowledge of the target node's network ID, i.e., its IP address; this is often unknown in practice. 
These challenges motivate the Peri algorithm in~\sref{sec:design}, which avoids all the above assumptions.

\section{Design}
\label{sec:design}
Optimizing network latency requires knowledge of network topology and/or node network addresses (\sref{sec:model}), which are unknown for typical P2P networks.
Instead, an agent is typically only aware of its own peers.
In this section, we present the \peri peering algorithm to account for this and achieve \textit{reduced} (if not necessarily optimal) latency.
\peri is a variant of the Perigee \cite{mao2020perigee} algorithm. 

\subsection{Perigee}
\label{sec:dml:perigee}
Perigee was introduced in \cite{mao2020perigee} as a decentralized algorithm for generating network topologies with reduced broadcast latency for transactions and blocks.
The Perigee algorithm is presented in \aref{alg:perigee}.
Lines that are highlighted in \textcolor{red}{red} are specific to \peri (\sref{sec:peri}).

\begin{Ualgorithm}
    \SetAlgoLined
    \SetKwFunction{Fpm}{\code{peer\_manager}()}
    \SetKwProg{Fn}{Run Thread}{:}{}
    \KwIn{Number of peers $K$ kept after each iteration, maximum peer count $N$, 
    length of period $T$, score function \textcolor{red}{$\phi$}}
    \KwResult{Peer set $P$ at each period $h = 1, 2, \cdots$}

    $P \leftarrow \emptyset$, \textcolor{red}{$B \leftarrow \emptyset$} \tcp*[r]{$B$ is a blocklist of evicted peers}
    
    \textcolor{red}{
    \Fn{\Fpm}{
        \While{\code{true}}{
            Hang and wait for next peer\;
            $v \leftarrow $ Random sampled node from $\Vc - B$\;
            \If{$|P| < N$}{
                $P \leftarrow P \cup \{v\}$ \tcp*[r]{add peer when a slot is available}
            }
        }
    }
    }

    \For{ $h \leftarrow 1, 2, \cdots $ }{
        Sleep $T$ \tcp*[r]{\code{peer\_manager} adds peers to $P$ when sleeping}
        $e \leftarrow N - K$ \tcp*[r]{$e$ is the number of peers to evict}
        Init score map $\Phi$\;
        \For{ $p \in P$ }{
            \If{ \textcolor{red}{\textbf{not}~~ \code{is\_excused}($p$) } }  { 
                $\Phi(p) \leftarrow \textcolor{red}{\phi(p)}$  \label{line:phi}\;
            } \textcolor{red}{ \Else { 
                $e \leftarrow e - 1$\;}
            }
        }
        \If{ $e > 0$ }{
            $P \leftarrow P \,- ~$\{$e$ keys with largest values in $\Phi$\}\;
            \textcolor{red}{$B \leftarrow B\, \cup ~$\{$e$ keys with largest values in $\Phi$\}}\;
        }
        \textbf{Output} $(h, P)$\;
    }
    
    \caption{\textbf{Perigee \cite{mao2020perigee}/ \textcolor{red}{Peri}.} \textcolor{red}{Red} text denotes parts that are specific to \peri. Note: 
    \code{is\_excused} is a predicate that is true when peer-delay information is insufficient to make a peering decision. 
    \textcolor{red}{$\phi(v)$} equals the average transaction-delivery delay of $v$ with respect to the best peer, as defined in Eqn.~\eqref{eqn:score}. It incorporates design choices 
    highlighted in \sref{sec:peri}.
    \vspace{-3ex}
    }
    \label{alg:perigee}
\end{Ualgorithm}

At a high level, Perigee requires every node in the network to assign each of its peers a latency score and periodically tears down connections to peers with high scores.
In our context, the latency score represents the latency with which transactions are delivered; we want this score to be as low as possible.
Over time, Perigee causes nodes to remain connected to low-latency peers,
while replacing other peers with random new ones.
Roughly speaking, \cite{mao2020perigee} shows that Perigee converges to a topology that is close to the optimal one, in the sense of minimizing global broadcast latency.


More precisely, Perigee divides time into periods.
Let $u$ denote a given node in the network and $M_v$ denote the set of all transactions received by $u$ in the current period from a current peer $v$. 
For a given transaction $m$, let $T(m)$ denote the time when $m$ is first received by $u$ from any of its peers  and $T_v(m)$ denote the time when $m$ is received by $u$ specifically from  $v$. 
We define $T_v(m) = \infty$ if $v$ did not deliver $m$ during the current period. 

In every period, each node $u$ evaluates a score function $\phi$ over each of its peers $v$, defined as
\begin{equation}
    \phi(v) \triangleq \sum_{m \in M_v}  \frac{1 }{ \abs{M_v}  } \min\left\{ T_v(m) - T(m),~~ \dmax\right\}.
    \label{eqn:score}
\end{equation}

\vskip-1ex
\noindent $\dmax$ is a parameter used by Perigee as an upper bound on measured latency differences. It bounds the influence of outliers on score-function computation.
For a node $u$, the score function $\phi(v)$ captures the average over all transactions $m$ of the difference in latency between delivery of $m$ by $v$ and that by the peer from which $u$ first received $m$. In other words, $\phi(v)$ may be viewed as the average slowdown imposed by $v$ with respect to the fastest delivery of transactions to $u$.


The procedure
\code{peer\_manager} is an asynchronous thread (or set of threads) that handles peer connections on the P2P network,
including accepting incoming peer requests, requesting nodes for connection and dropping peers.
Ideally, it can randomly sample nodes from the entire network, gradually add peers when the peer count is under the maximum,
and keep connections with specific nodes (targets).
Hence, while the main thread sleeps, \code{peer\_manager} expands the peer set $P$ until it reaches its maximum size $N$.

\subsection{\peri}
\label{sec:peri}
Although Perigee was designed to be deployed by \emph{all} network nodes to improve broadcast latency, we observe that the same ideas can be applied by a single agent to improve their observed direct and triangular latency. 
We next describe \peri,  a slight modification of Perigee enabling an agent to control their direct and triangular latency. 
Although \peri does not functionally differ from Perigee, we give them different names to differentiate their usage and implementation choices. 
Again, the steps unique to \peri are highlighted in \textcolor{red}{Red} in \aref{alg:perigee}.
The main differences  are:
\begin{enumerate}[leftmargin=*]
    \item \textbf{Goal:} \peri is meant to be applied by a single node to advantage it over other nodes, whereas Perigee was proposed as a protocol to optimize systemic network performance. 
    \item \textbf{Relevant transactions:} In Perigee, nodes measure the latency of all received transactions. In \peri, the score function $\phi(v)$ enforces that only \emph{relevant transactions} participate in scoring peers. 
    In Peri, for \latg{} reduction, all  transactions are considered relevant, while for \latt{} reduction, only transactions made by the target are relevant. 
    \item \textbf{Handling silent peers:} Particularly when optimizing targeted latency, the function $\phi$ may be undefined for some peers in some periods.
    For example, if a peer $p$ is connected near the end of a given period,
    there will not be sufficient data to compare $p$ with other peers,
    which means $\phi(p)$ may be undefined.
    Instead of evicting $p$ in such cases and possibly losing a good peer,
    we forego eviction of $p$. In \aref{alg:perigee}, the predicate \code{is\_excused}($v$) is true if node $v$ should be excluded from eviction for the current period.
    \item \textbf{Blocklists.} The Perigee \cite{mao2020perigee} algorithm advocates for selecting a new set of peers at random. 
    However, this increases the likelihood of a peer tearing down connections, then connecting to the same node(s) shortly thereafter, particularly since some cryptocurrency clients (including \texttt{geth}) favor previously-visited nodes during peer selection \cite{heilman2015eclipse,henningsen2019eclipsing}. 
    To handle this, in \peri we use \emph{blocklists}: if a node tears down a connection to peer $v$ in a \peri period, it refuses to re-connect to $v$ in future periods. In \aref{alg:perigee}, we maintain blocklist $B$ for this purpose.
    \item \textbf{Sampling relevant transactions:}  
    We cannot relay relevant transactions; otherwise, the ``late'' peers who do not deliver a relevant transaction first to our node will \emph{never} deliver it to our node (\sref{sec:background}), thus removing them from \peri's latency comparison.
    If all transactions are relevant, our node will hence act as a black hole, and may impact the P2P network. 
    We avoid this by sampling 1/4 of all relevant transactions for global latency measurement,
    by redefining relevant transactions  as those with hashes divisible by 4 when computing $\phi(\cdot)$ in Line \ref{line:phi} of Alg. \ref{alg:perigee}.
    %
 
\end{enumerate}
\section{Direct Latency Evaluation}
\label{sec:dml}


In \sref{sec:dml:global} and \sref{sec:dml:targeted}, we show the practical latency reduction effects of \peri with experiments on the Ethereum P2P main network (mainnet) and Rinkeby test network (testnet).

\subsection{Evaluation: Direct Global Latency}
\label{sec:dml:global}
\subsubsection{Methods}
\label{sec:baselines}
We evaluate four approaches.
\begin{enumerate}[leftmargin=*]
    \item \textbf{Baseline.} Our experimental control node uses the default settings of the Go-Ethereum client, version 1.10.16-unstable \cite{goethereum}\footnote{The customized client source code can be found at \redact{\href{https://github.com/WeizhaoT/geth_peri}{\color{blue}https://github.com/WeizhaoT/geth\_peri}}.}. This was the latest version when we started the experiments. 
    \item \textbf{BloXroute.} We compare against a centralized, private relay network, using bloXroute as a representative example. We use the bloXroute Professional Plan \cite{BloXroute}; at the time our experiments were run, this plan cost \$300 per month.
    \footnote{\resub{
        bloXroute also offers more expensive and powerful plans, which we did not test due to financial constraints. 
    }}
    We ran the bloXroute gateway locally to avoid incurring additional latency (\sref{sec:related}). 
    \item \textbf{\peri}. We modify the Go-Ethereum client \cite{goethereum} to implement the \peri algorithm for peer selection. We set the period to 20 minutes, and replace at most 25 peers every period.
    \item \textbf{Hybrid.} We implement a hybrid method that combines bloXroute and \peri by applying \peri to a node with access to a bloXroute relay.
          For correctness, we ensure that the gateway connecting to the relay, which acts as a peer of the node, cannot be removed by the \peri algorithm.
\end{enumerate}



\subsubsection{Experimental Setup.}
We establish 4 EC2 instances in the us-east-1 AWS data center, where a public bloXroute relay is located.
On each instance, we deploy a full node on the Ethereum P2P main network (the mainnet), which is implemented by a customized Go-Ethereum (geth) client.
Each node has at most 50 peers, which is the default setting of Go-Ethereum.
For a node running \peri or Hybrid, we set the proportion of outbound peers (peers dialed by the node itself) to 80\% so that the node actively searches for new peers.
For a node running other baselines, we keep the proportion at 33\%, which is also default for Go-Ethereum.

We ran 63 experiments from Feb 18, 2022 to March 16, 2022, each following the procedure below.
First, we assign each latency reduction method (bloXroute, \peri, Hybrid and Baseline) exclusively to a single node, so that all four nodes use different comparison methods.
Then, we launch the nodes simultaneously.
When a packet arrives, the node checks if the packet contains a full relevant transaction or its hash; if so, it records the timestamp.
At the end of the experiment, we stop the nodes and collect the arrival timestamps of transactions from their host instances.

\paragraph{Bias reduction.}
We take the following steps to control for systematic bias in our experiments.
We prohibit the 4 measurement nodes from adding each other as peers.
We reset all the enode IDs (unique identifiers of the Ethereum nodes including IP address) before each experiment. This prevents the other nodes from remembering our nodes' IP addresses and making peering decisions based on activity from previous experiments.
We record the clock differences with periodic NTP queries between each pair of hosts to fix systematic errors in the timestamps recorded locally by the hosts.
Despite being the same type of AWS instance, the host machines may also introduce biases because they may  provide different runtime environments for the Ethereum node program.
To eliminate these biases, we rotate the assignment of latency reduction methods after every experiment, so that for every successive 4 experiments, each node is assigned each method exactly once.
We attempt to control for temporal biases due to diurnal transaction traffic patterns by running each experiment every 8 hours, so that the assignment of methods to nodes  rotates 3 times a day.
Because the number of possible method assignments is 4, a co-prime of 3, each node experiences every combination of latency reduction methods and time-of-day.
We did not control for contention (e.g., of network bandwidth, computational resources) among EC2 instances by running experiments on dedicated hardware due to financial constraints.

\subsubsection{Results}
\begin{figure}
    \includegraphics[width=.75\linewidth]{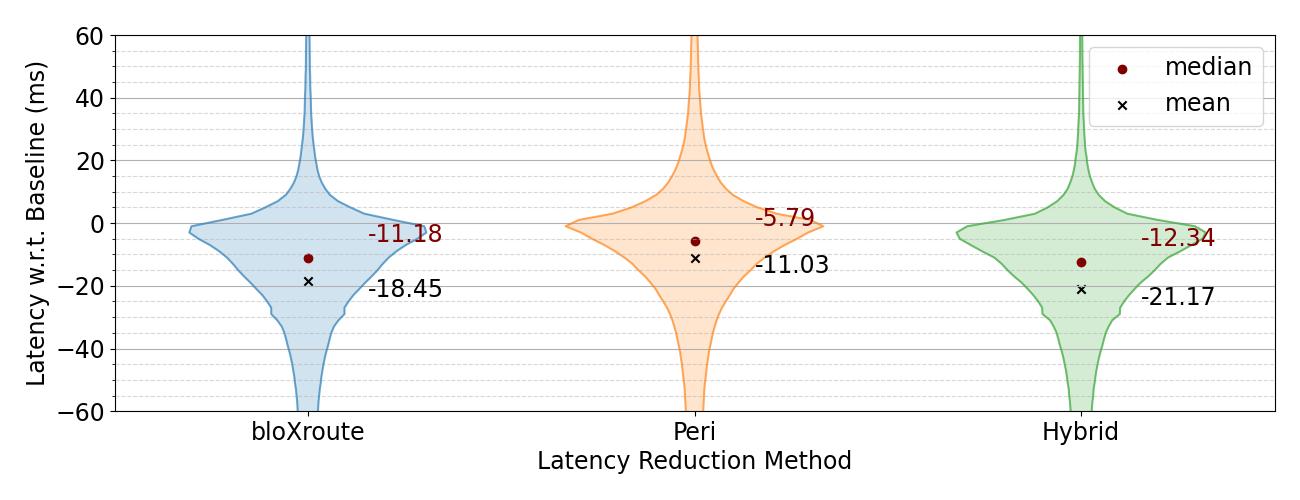}
    \caption{PDFs of global latency distributions (difference in ms from Baseline latency).
    \vspace{-4ex}
    }
    \label{fig:latg}
\end{figure}

Each node is allowed to warm up for 2.5 hours; after this, we collect all transactions that are received by each of the nodes for 3.5 hours. 
\resub{
    Although it is infeasible to measure the transaction propagation time directly---as this would require sender timestamps---the time difference of arrival allows us to measure the ability of a method to \emph{reduce} propagation time. 
}
For each transaction $m$ and each node $y$ with a latency reduction method other than Baseline,
we compute the difference between the timestamp of $m$ at $y$ and the timestamp at the baseline node $b$,
which is effectively a sample of random variable $\Lambda(y) - \Lambda(b)$.
The smaller the time difference (i.e., the more negative), the earlier $y$ delivers $m$, and the more effective the latency reduction method is.
We gather the latency differences over 63 experiments, and plot their distributions for each node in \fref{fig:latg}.
In total, we analyzed the latencies of 6,228,520 transactions.

\begin{finding}
    \peri alone is at least half as effective as bloXroute private networks in reducing average \latg{}.
    The Hybrid node (\peri with bloXroute) achieves an additional 15\% improvement in latency reduction over bloXroute alone.
\end{finding}


In the figure, all the latency differences are distributed with negative means and medians, showing an effective latency reduction.
We find these results to be statistically significant; our statistical tests are detailed in App. \ref{app:stats}.
BloXroute reduced this latency by 18.45 milliseconds on average. 
\resub{
    In comparison, \peri reduced it by 11.09 milliseconds. 
    Although \peri's ability to reduce latency is expected, 
    it is perhaps unexpected that \peri is more than half as effective as bloXroute (and cost-free). 
}
On the other hand, by exploiting access to bloXroute services, Hybrid nodes can achieve an \textit{additional} 15\% reduction in latency.
This suggests the potential of peer-selection algorithms to further boost latency reduction techniques over private relay networks.

\cam{
Note that these experiments naturally experienced network dynamics (e.g., traffic congestion) over months on the live Ethereum P2P network, suggesting that the results are robust to realistic conditions. On the other hand, it is difficult to empirically determine how network dynamics (cross traffic, BGP, etc.) exactly affect our results, because we cannot control or even measure them comprehensively in the network. As a compromise, we study the effects of changes in local network conditions, including the peer count and bandwidths of the attacker.
}
\sigm{
Roughly, we found that the higher the adversary's peer count limit, the lower the advantage afforded by Peri; above a peer count of 100, the benefits of Peri appear to plateau, with median reductions between 2-5 ms (App. \ref{sec:eval_pc}). Similarly, we find that if the attacker has a low-bandwidth connection to the network, the benefits of Peri are significantly amplified. For example, if we throttle the adversary's bandwidth to 1.2 Mbps, the median latency reduction of Peri compared to the baselins is 37 ms---$7\times$ larger than the reduction in Fig. \ref{fig:latg}, which uses a 10 Gpbs link  (App. \ref{sec:eval_bw}). Overall, these results suggest that \textbf{the benefits of strategic latency reduction are most significant for nodes with comparatively low network resources.}}

\subsection{Evaluation: Direct Targeted Latency}
\label{sec:dml:targeted}


Experimentally evaluating targeted latency reduction techniques on the Ethereum mainnet can be costly and time-consuming.
This is because to evaluate latency distributions, we need to observe many transactions with a known ground truth IP address.
A natural approach is to generate our own transactions from a single node and measure their latency; unfortunately, nodes in the P2P network do not forward transactions that are invalid or unlikely to be mined due to low gas fees.
Hence, we would need to create valid transactions.
For example, at the time of writing this paper, the recommended base fee of a single transaction was about \$2.36 per transaction \cite{ethgasstation}. Collecting even a fraction of the 6+ million transactions analyzed in \sref{sec:dml:global} would be prohibitively costly.

We ran limited experiments on \latt{}, where the setups and results are shown in details in Appendix \ref{sec:exp-targeted-mainnet}. 
The findings of these experiments are summarized below. 

\begin{finding}
    \peri reduces \latt{} by 14\% of the end-to-end delay, which is as effective as bloXroute and Hybrid. 
\end{finding}


We did not collect a sufficient amount of data for \peri and Hybrid to show a statistically significant ability to connect directly to a victim and learn its IP address.
We attribute this to the low transaction frequency and large size of the mainnet. 
The testnet experiments we describe next, however, \textit{did} result in frequent victim discovery of this kind---provided a sufficiently large number of Peri periods or a high frequency of victim transactions.

\subsubsection{Targeted Latency on Ethereum testnets}
\label{sec:targeted-latency-testnet}
Due to the financial cost of latency reduction experiments on the Ethereum mainnet, we also performed experiments on the Rinkeby testnet.
\footnote{Though the Ropsten testnet is generally thought to be the test network that most faithfully emulates the Ethereum mainnet \cite{testnets}, 
its behavior was too erratic for our experiments during our period of study. 
This included high variance in our ability to connect to peers from any machine and constant deep chain reorganizations, 
causing lags in client synchronization.} 
Our goal in these experiments is to simulate a highly active victim in the network and compare the \peri algorithm's ability to find and connect to the victim against a baseline default client.

In these experiments we run a victim client on an EC2 instance in the ap-northeast-1 location and the \peri and Baseline clients in the us-east-2 location. 
Note that we cannot evaluate bloXroute or Hybrid on the test network, because bloXroute does not support test network traffic. 
Due to our observations of the Rinkeby network's smaller size, and to emulate a long-running victim whose peer slots are frequently full, we set the peer cap of the victim to 25 peers. 
As in previous experiments, the Baseline node is running the default geth client with peer cap of 50. 
The \peri node is also running with a peer cap of 50 but with the \peri period shortened to 2 minutes. 

\begin{minipage}{.96\linewidth}
\begin{minipage}[b]{0.35\linewidth}
    \centering
    \begin{tabular}{p{3.5em}<{\centering}p{8em}<{\centering}}
        \hline
        \textbf{Method} & \textbf{Successful Victim Connections} \\ 
         \hline
        \peri & 47/70 = 67.1\% \\
        Baseline & 6/70 = 8.6\% \\ 
        \hline
    \end{tabular}
    \captionof{table}{Rate of connection to victim node in Rinkeby testnet experiments.}
    \label{tab:connections}
\end{minipage}
\hfill
\begin{minipage}[b]{0.58\linewidth}
    \centering
    \includegraphics[width=\linewidth]{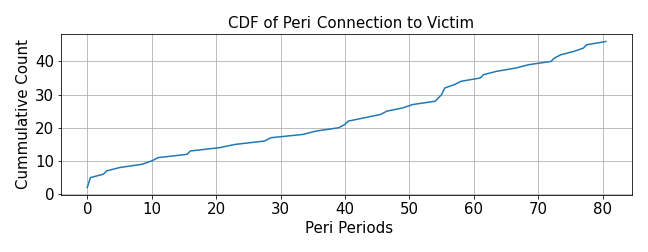}
    \captionof{figure}{CDF of number of \peri periods before finding victim. Each \peri period lasts 2 minutes.}
    \label{fig:latt_testnet_positioning}
\end{minipage}
\end{minipage}

\vskip1ex

In each experiment, we start the victim client first and give it a 30 minute warm-up period to ensure its peer slots are filled. We then start the \peri and Baseline nodes and begin transmitting transactions from multiple accounts on the victim node.
We set the frequency to $3$ transactions per minute to control the variance of timestamps and \peri scores.
We then run the experiment for just under 3 hours, enabling us to run 7 experiments per day and  alleviate time-of-day biases.
\sigm{After each experiment, we say the \peri node found the victim if the victim's Peri score is best (lowest) among the currently-connected peers. 
Since the baseline does not compute Peri scores, we say it found the victim if the baseline node ever connected to the victim.
}
As with the global latency experiments in \sref{sec:dml:global}, we alternate which us-east-2 machine is running \peri v.s. Baseline across runs in order to mitigate potential machine-specific biases. 
We ran these experiments from April 16 - 26, 2022, for a total of 70 runs.

\begin{finding}
    In testnet experiments, \peri is able to identify the IP address and connect to a target (victim) node with a frequency more than 7$\times$ that of a Baseline node.
\end{finding}

\begin{wrapfigure}{l}{0.5\linewidth}
        \includegraphics[width=\linewidth]{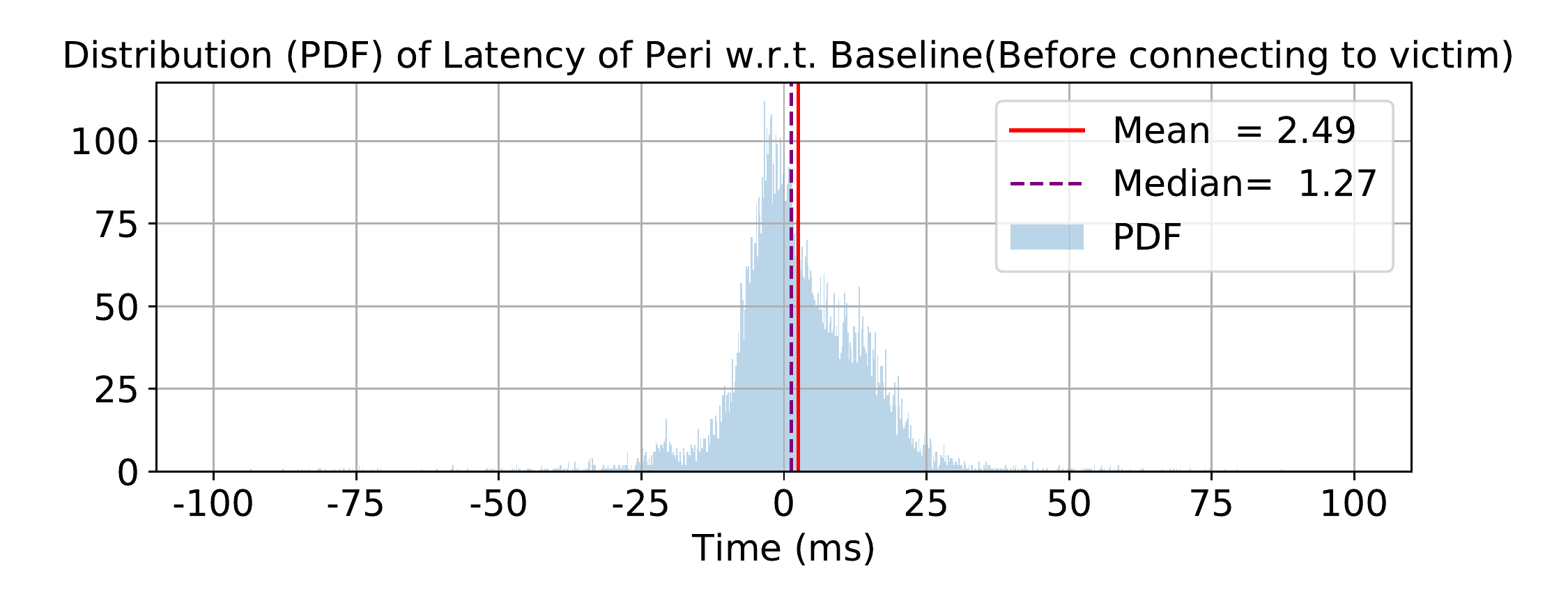}
    \\
        \includegraphics[width=\linewidth]{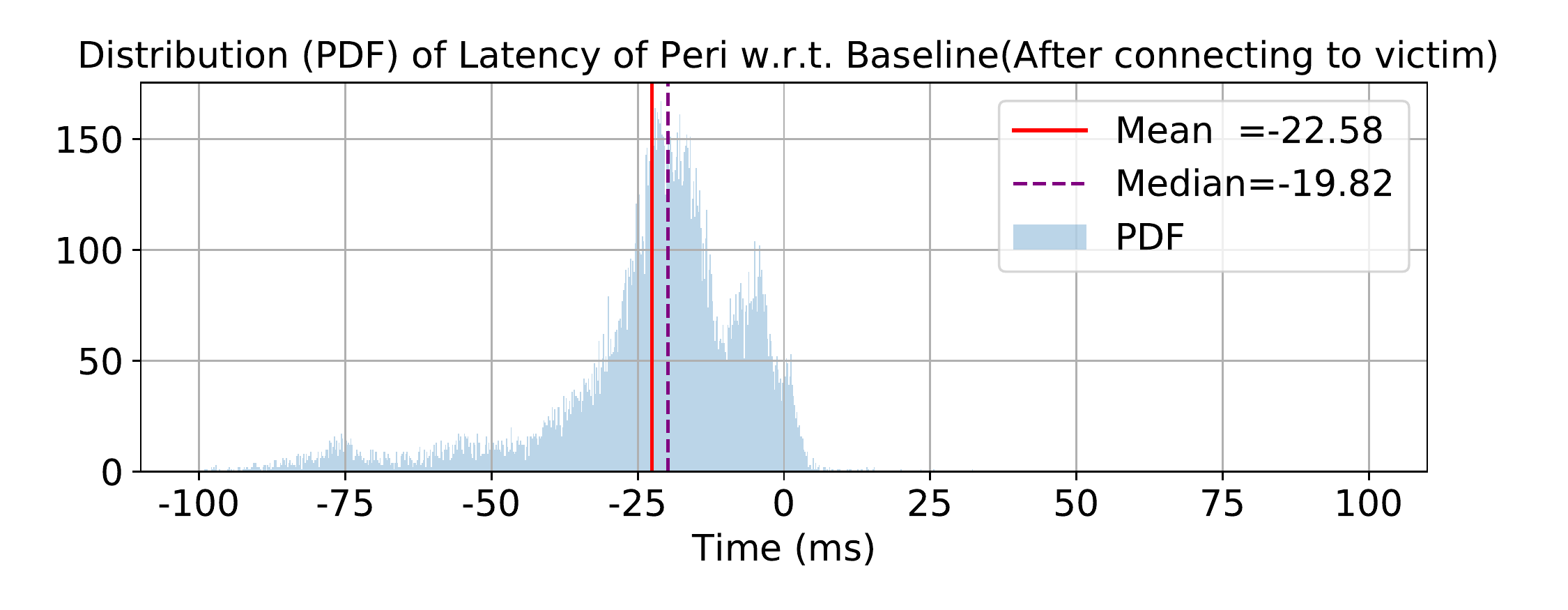}
    \caption{PDFs of distributions of targeted latency for runs when \peri found the victim split between before either found the victim, and after \peri finds the victim. The $x$ axis represents the difference of timestamp at the \peri node and at the baseline node for each single transaction. 
    }
    \label{fig:latt_testnet}
\end{wrapfigure}

The number of connections established by the \peri and Baseline clients to our victim are listed in Table \ref{tab:connections}.
We find that \peri gives a notable advantage when connecting to the victim node. 
\fref{fig:latt_testnet_positioning} shows a CDF of the number of \peri periods until \peri connected to the victim, with an mean/median of 39/45 periods ($\sim 1.3/1.5$ hrs). 
\fref{fig:latt_testnet} shows the delay PDF for the runs when \peri finds the victim for all transactions before and after \peri connects to the victim. 
We observe a mean latency advantage of 22ms over the Baseline client once \peri connects to the victim. 
This is 30\% of the end-to-end delay between the victim and both the \peri and Baseline clients. 
Unlike our mainnet experiments, \peri has no significant latency advantage before it connects to the victim. 
We attribute the difficulty of establishing a network advantage prior to direct connection to the victim to the Rinkeby network's smaller size (1.8K unique IPs we came across over the period of our study compared to 13.6K in the mainnet). 
Hence, there may be relatively few peers in geographical proximity to the the Tokyo data center compared to our mainnet victim node in Germany.
There are significant differences in size and topology between the Rinkeby network and the Ethereum mainnet, and it is unclear how our experiments extrapolate to the mainnet. 

\cam{
Though it is interesting to extend our results to different settings (different peer count, etc.), we encounter limitations for both the testnet and the mainnet. 
On the testnet, there are not enough active nodes to support a peer count of even 50, and the latency data is much noisier than on the mainnet. 
On the mainnet, the victim node from \redact{Chainlink} is no longer available. 
}


    




\section{Triangular Latency Evaluation}
\label{sec:triangular-latency}
A front-runner may be interested in targeted and/or global triangular latency. 
As with targeted latency, the optimal strategy for an agent to reduce triangular targeted latency is to connect directly to the target nodes (\sref{sec:triangular-metrics}). 
Since this procedure is identical to the experiments in \sref{sec:dml:targeted}, we do not run additional experiments to demonstrate its feasibility.

The more complicated question is how to optimize triangular global latency. 
Recall from~\sref{sec:triangular-metrics} that we study triangular global latency via a proxy metric, which we call \emph{adversarial advantage} $A_{\Nc}(U)$, where $\Nc$ denotes the network and $U$ denotes the set of strategic or adversarial agents.  
In the remainder of the section, we first show that directly optimizing adversarial advantage is computationally infeasible even if the agent knows the entire network topology (\sref{sec:triangular-hardness}).
However, we also show through simulation that 
a variant of \peri can be used to outperform baselines (\sref{sec:triangular-approximations}).\footnote{\textit{Note on evaluation:} Adversarial advantage is more difficult to evaluate empirically than direct latency. 
For a single source-destination pair, we would need to observe transactions from a known source (e.g., a front-running victim), which end up at a known target destination (e.g., an auction platform \cite{FlashbotsDashboard}); we would also need to verify that our agent node can reach the target node \emph{before} the victim's transaction reaches the target destination. 
Setting up mainnet nodes to measure this would be twice as costly as evaluating direct targeted latency, which we deemed infeasible in \sref{sec:dml:targeted}.
Measuring adversarial advantage globally would further require visibility into every pair of nodes in the network. 
We therefore evaluate triangular latency reduction theoretically and in simulation.}


\subsection{Hardness of Optimizing Adversarial Advantage}
\label{sec:triangular-hardness}

Generally, we are interested in the following problem.
%
\begin{problem*}
    Given a network $\Nc$, sets of sources and destinations $\Sc, \Tc$, and budget $k$ (number of edges the agent can add), we want to maximize $A_{\Nc}(U)$ (Definition \ref{def:advantage}) subject to $|U | \le k$, where $U$ is the set of peers to which the agent node connects.
    \label{prob:max_adv}
\end{problem*}

\begin{theorem}
    Problem \ref{prob:max_adv} is NP-hard.
    \label{thm:hardness}
\end{theorem}
(Proof in Appendix \ref{sec:app-proofs-hardness})
The proof follows from a reduction from the set cover problem. 
Not only is solving this problem optimally NP-hard, we next show that it is not possible to \emph{approximate} the optimal solution with a greedy algorithm.
A natural greedy algorithm that solves the advantage maximization problem is presented in Algorithm~\ref{alg:greedy}. %
\footnote{It includes a ``bootstrapping'' step in which two nodes are initially added to the set $U$ of agent peers, as no advantage is obtainable without at least two such nodes.}
It is \emph{unable} to approximately solve Problem \ref{prob:max_adv}. 

\begin{algorithm}[htbp]
    \KwIn{Number of peers $k \ge 2$, graph $\Nc$}
    \KwOut{Set of peers $U$}

    $U \leftarrow \argmax_{\{x, y\} : (x, y) \in \Vc^2} A_{\Nc}(\{x, y\}) $ \tcp*[r]{bootstrapping} 

    \For{$j \in \{3, 4, \cdots, k\}$}{
        $\bar z \leftarrow \argmax_{z: z \in \Vc - U} A_{\Nc}(U \cup \{z\}) $\;
        $U \leftarrow U \cup \{\bar z\}$\;
    }

    \Return $U$\;
    \caption{Greedy Algorithm for Maximizing $A_{\Nc}$. }
    \label{alg:greedy}
\end{algorithm}




\begin{proposition}
    The output of Alg. \ref{alg:greedy} is \emph{not} an $\alpha$-approximate solution to Problem \ref{prob:max_adv} for any $\alpha > 0$. 
    \label{prop:greedy_non_approx}
\end{proposition}

(Proof in Appendix \ref{apdx:greedy_non_approx})
We show this by constructing a counterexample for which the greedy algorithm achieves an adversarial advantage whose suboptimality gap grows arbitrarily close to 1 as the problem scales up.
Whether a polynomial-time approximation algorithm exists for maximization of $A_{\Nc}$ is an open problem.

\subsection{Approximations of Advantage Maximization}
\label{sec:triangular-approximations}

Although the greedy algorithm (Alg. \ref{alg:greedy}) is provably suboptimal for maximizing adversarial advantage,  we observe in simulation that for small, random network topologies (up to 20 nodes), it attains a near-optimal adversarial advantage (results omitted for space). 
We also show in this section that the greedy strategy achieves a much higher advantage (2-4$\times$) than a natural baseline approach of randomly adding edges. 
Hence, the greedy strategy may perform well in practice. 
However, the greedy strategy \emph{requires knowledge of the entire network topology}. 

We next evaluate the feasibility of local methods (i.e., \peri) for approaching the greedy strategy.\footnote{The simulator code can be found at  \redact{\href{https://github.com/WeizhaoT/Triangular-Latency-Simulator}{\color{blue}https://github.com/WeizhaoT/Triangular-Latency-Simulator}}.}

\paragraph{Graph topology.}
We consider four models of random graph topologies: {\ergraph{}, }random regular graphs, \bagraph{} (scale free) graphs, and Watts–Strogatz (small world) graphs---each with 300 nodes.
\sigm{In addition, 
we study a snapshot of the Bitcoin P2P network
\cite{miller2015discovering} and the Lightning Network (LN) \cite{lngossip} (sampling details in App. \ref{sec:sim_cent_all}).
}
The Ethereum P2P network has been shown to exhibit properties of both small world and scale-free networks \cite{wang2021ethna}.
The average degree is set to 9 for the \ergraph{} model to ensure connectivity.
The average degree is set to 4 for the other 3 models to make sure the average distance between nodes is high enough for shortcuts to exist. 
For each model, we sample 25 different graph instances. 
To model the observed existence of hubs in cryptocurrency P2P networks \cite{miller2015discovering,delgado2019txprobe,eisenbarth2021open,deshpande2018btcmap}, we introduce 20 new hub nodes, each connecting to 30 nodes randomly sampled from the original graph.

On each instance, the set of sources $\Sc$ is equal to the set of all the nodes $\Vc$, and the set of destinations $\Tc$ is a random subset of $\Vc$ with $|\Tc|=0.1 \cdot |\Vc|$. Note that $\Tc \subset \Sc$. 
We let the static front-running penalty $\tau = 0$, a minimum path advantage, and assign a unit weight to all the edges.

\paragraph{\peri Modification.}
To make \peri optimize adversarial advantage, we alter the scoring function and the peer-selection criterion; 
this allows us to simulate the (modified) \peri algorithm without simulating individual messages. 
First, we make the following simplifying assumptions:
    (a) Peer replacement is completed immediately at the beginning of a new \peri period, and no peer is added or dropped during the period;
    (b) The full set $M$ of messages sent during the current \peri period is delivered to the adversarial agent during the \peri period by every peer; 
    (c) The end-to-end delay of each message is proportional to the distance between its source and destination; 
    \label{item:propdelay}
    (d) Message sources are uniformly distributed over the network \sigm{and message destinations  send messages at the same frequency; and
    (e) The agent can tell whether the source of any message belongs to set $\Tc$. }

Let $d_v(m)$ denote the distance from \sigm{$s(m)$,} the source of $m$, to peer $v$ of the agent node $a$,
and $S(m)$ the time when message $m$ departs from the source. 
\sigm{
As in \eqref{eqn:score}, we define $T_v(m)$ as the time when $m$ is received from peer $v$, and $T(m)$ as the time when $m$ is first received from the fastest peer.
}
Recall that we assume $d_{\Nc'}(v, a) = 0$ for any peer $v$ of $a$  
in~\sref{sec:triangular-metrics}, so for such a $v$, $d_v(m)$ equals the distance from the source of $m$ to $a$. 
We assume the agent assigns a weight in the score function $\lambda_{s(m)}$ by the source $s(m)$. 
The score function $\phi(v)$ for peer $v$ can be transformed with constants $C_1, C_2, C_3$ with respect to $v$ for a set topology.
\begin{align}
    \phi(v) 
             = \sum_{m \in M}  \frac{T_v(m) - S(m)}{|M| \lambda_{s(m)}^{-1}} + \frac{S(m) - T(m)}{|M| \lambda_{s(m)}^{-1}} 
             = C_1 \sum_{m \in M} \lambda_{s(m)} d_v(m) + C_2 = C_3 \sum_{s \in \Sc}\lambda_{s} d_{\Nc}(s, v) + C_2. \notag
\end{align}

Since \peri's choice of peers to drop is invariant to $C_2$ or $C_3$,
we can further simplify the score:
$
    \phi(v) = \sum_{s \in \Sc} \lambda_s d_{\Nc}(s, v). 
$
\sigm{
Recall that $\Tc \subset \Sc$, which implies that by properly controlling $\lambda_s$, we can let the agent pay equal attention to approaching sources and destinations. Eventually,\
\begin{equation}
    \phi(v) = \frac{1}{|\Sc|} \sum_{s \in \Sc}  d_{\Nc}(s, v) + \frac{1}{|\Tc|} \sum_{t \in \Tc}  d_{\Nc}(t, v). \label{eqn:score_simple_final}
\end{equation}
}

For a given peer budget $k$, we replace 
$r = \left\lceil\frac{k}{3}\right\rceil$ 
peers and keep $k$ peers. 
We only consider the kept peers for evaluation of the advantage metric.
In total we execute 800 \peri periods.

\begin{figure}
    \centering
    \includegraphics[width=\linewidth]{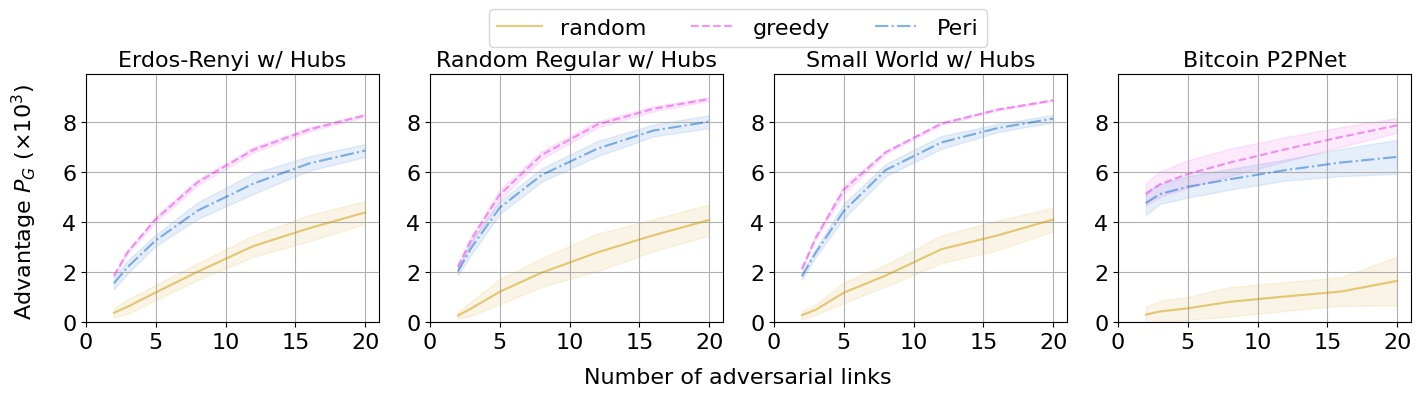}
    \caption{Advantage-Peer-count curves on hub-enriched graph models. The mean of the curves are shown by solid lines, and the standard deviations are shown by the transparent color zones centered at the mean. 
    }
    \vspace{-0.4cm}
    \label{fig:sim_cent_top}
\end{figure}

\paragraph{Results.}
We sweep the peer count budget $|U| = k$ of the agent over 7 values ranging from 2 to 20.
For each maximum peer count, we obtain a resulting peer set $U^*$ with advantage $A_{\Nc}(U^*)$ using the greedy algorithm, the \peri algorithm, and random sampling, in which each agent chooses to peer with nodes selected uniformly at random.
The performance of each method is represented by its advantage-peer-count ($A_{\Nc}$-$|U|$) curve.
For each graph model, we plot the mean and standard deviation of the curves over 25 different hub-enriched graph instances in \fref{fig:sim_cent_top}.

\begin{finding}
\peri is competitive with the greedy algorithm for maximizing adversarial advantage when the underlying network has many hubs, or nodes of high degree.\footnote{\sigm{Precisely, we define a hub as a node whose degree is at least 10\% the total number of nodes in the graph. }}
\end{finding}

Regardless of original topology, the existence of hub nodes enables the \peri algorithm to place shortcuts almost as effectively as the greedy algorithm.
We notice that \peri worked significantly better on models with more hubs such as scale-free than the other topologies. 
It is also notable that over 70\% of the resulting peers of the \peri algorithm are hubs. 
\sigm{
    Our results on Bitcoin and the Lightning Network (App. \ref{sec:sim_cent_all}, \ref{sec:sim_orig}), however, show, that in practice, the gap between Peri and the Greedy baseline grows with the number of adversarial links. 
    Peri appears to be too aggressive in selecting hubs, which are often clustered in real-world networks; hence connecting to multiple hubs gives less advantage.
    Nontheless, Peri still outperforms the random baseline by $3\times$ - $5\times$.
}

\section{Impossibility of \Fairness}
\label{sec:frl:impossibilty}

\peri and its variants, e.g., Hybrid, can significantly improve direct and triangular latency.
A natural question is whether this is because most nodes are currently running a poor peer selection strategy (Baseline). 
If \emph{every} node were running \peri or Hybrid, would it still be possible for an agent to strategically improve their targeted latency?
We show that no matter what peering strategy nodes use, if the P2P network has some natural churn and target node(s) are active, a strategic agent can \emph{always} manipulate their targeted latency.


\subsection{Model}
We start with some additional modeling assumptions regarding the temporal dynamics of our network.
Consider a time-varying network $\Nc(t) = (\Vc(t), \Ec(t))$ where $\Vc(t)$ denotes the set of nodes and $\Ec(t)$ denotes the set of undirected edges of $G$ at time $t$.
In the network, any party can spawn a set $V'$ of  nodes and add them to $\Vc(t)$. ($\Vc(t^+) = \Vc(t) \cup V'$, where $t^+$ denotes the time infinitesimally after $t$).
We assume there is an upper bound $\bar V$ on the total number of nodes in the network; that is, $\Vc(t)\leq \bar V$ for all $t\ge 0$, due to the limited address space for nodes.

If a node $u$ knows the network (Class 1) ID of another node $v$, 
$u$ can attempt to add an edge, or a peer connection, $(u, v)$ at time $t$ (that is, $\Ec(t^+) = \Ec(t) \cup \{(u, v)\}$) to the network.
This peering attempt will succeed unless all of $v$'s peer connections are full.
Nodes have an upper bound of $h$ on their total number of peer connections:
$
  \deg (u) \leq h ~~\forall u\in \Vc(t),~~\forall t\ge 0.
$
Among these, each node has at least $F \in (0, h]$ \textbf{dynamic peer slots}. 
A connection from $u$ to $v$ that occupies one of either $u$'s or $v$'s dynamic peer slots is called \emph{dynamic} and will be periodically torn down (see Definition \ref{def:p2pnet}).
Additionally, a dynamic peer slot is permissionless and is filled first-come-first-serve (FCFS). 

%

\resub{
    In practice, permissionless network nodes find each other via queries to distributed databases of network IDs in a process called peer discovery \cite{wang2019survey}.
    We model such peering databases as an oracle that responds to peer queries from any node. 
    In response to a query, the oracle independently draws the network ID of a node in the network from some (unknown) probability distribution,
    where each node may be drawn with a non-zero probability.
}
Specifically, there exists a universal constant $q > 0$ where the probability of drawing an arbitrary node is lower-bounded by $q$.
This is based on the assumption that the number of nodes is upper bounded.
We assume the oracle can process a fixed number of queries per unit time, so each query (across all nodes) takes constant time.
Upon processing a query, the oracle responds with the network ID of an existing node.

Each edge $(u, v) \in \Ec(t)$ has an associated \emph{link distance} $w(u, v)$, which operationally represents the end-to-end latency from $u$ to $v$.
We assume
\sigm{edge} latency $w(u, v)$ is \sigm{affected by the physical length of the path $u \to v$ on the Internet (i.e., including routing) and processing delays at the sender or recipient}.
Hence, we assume $w(u, v)$ is a constant over time and
satisfies the triangle inequality:
\begin{align}
    w(u, v) \le w(u, z) + w(z, v), \quad \forall u, v, z \in \Vc(t), t \ge 0. \label{eqn:triangle}
\end{align}
This is distinct from graph distance $d_{\Nc}(u, v)$ introduced in \sref{sec:model}.
\sigm{The triangle inequality often does not hold over the Internet \cite{lumezanu2009triangle}. 
However, we conjecture that triangle inequality violations may be less common in cryptocurrency P2P networks,
since end-to-end latency is significantly impacted by processing delays at router nodes, which scales with the number of hops in a route. 
}




\paragraph{Transaction dissemination.}
A node can broadcast an arbitrary message (transaction) at an arbitrary time through the entire network.
When a message is sent from $u$ to its neighbor $v$ at time $t$, $v$ will receive the message at time $t + w(u, v)$,
where $w(u, v)$ is the link distance between $u$ and $v$.
If $v$ is not adversarial, it will immediately forward the message to each of its neighbors.\footnote{This is a special case of randomized flooding protocols like diffusion \cite{fanti2017deanonymization}.}
We assume the P2P network is connected at all times $t \ge 0$.
We additionally assume that the latency between any pair of agent nodes is negligible, such that an agent connecting to multiple targets from multiple agent nodes, is equivalent to these targets peering with one node.

\paragraph{Liveness.}
In our analysis, we will assume that nodes are sufficiently active in terms of producing transactions and that the network experiences some amount of peer churn. 
We make both of these assumptions precise with the following definition.

\begin{definition}
    \label{def:p2pnet}
    A $(\lambda,\nu)$-\name{}  network has the following properties:


    \begin{enumerate}[label=(\Roman*),leftmargin=*]





        \item
              \label{aspt:dynamic}
              All {dynamic connections} have a random duration  $\text{Exp}(\lambda)$.



        \item For each node $u \in \Sc$, where $\Sc$ denotes a set of target nodes, we let $\{M_u(t),~ t \in [0, \infty)\}$ denote the counting process of messages that are generated and broadcast by $u$.
                  {$M_u(t)$ is a Poisson process with rate $\nu$.}
              %

              \label{aspt:activity}
    \end{enumerate}

\end{definition}

\subsection{Optimization of Targeted Latencies}
\label{sec:unfair-direct-targeted}

\subsubsection{Inferring Network ID}
\label{sec:unfair-infer}
We argue in \sref{sec:model} that we can optimize targeted latency (both direct and triangular) by connecting directly to the target node(s). 
However, in practice, an agent typically only knows the targets' logical (Class 2) IDs (e.g., public key of wallet), whereas it needs their network (Class 1) IDs (e.g., IP address) to connect. 
In the following, we show when an adversarial agent can learn the network ID(s) of one or more targets. 
Our first result states that in a \name{} network (Def. \ref{def:p2pnet}), an agent can uncover the network ID(s) of a set of target nodes with  probability $1-\epsilon$ given only their logical IDs in time quasilinear in $\epsilon^{-1}$ and quadratic in the number of target nodes.

\begin{theorem}
    Consider a \name{} network $\Nc(t) = (\Vc(t), \Ec(t))$.
    Let $\Ac$ denote an adversarial agent capable of identifying the logical ID of the sender of any message it received from the network.
    Given a set of target nodes $U$,
    with probability at least $1 - \epsilon$ for any $\epsilon \in (0, 1)$,
    it takes the agent $O(|U|^2 \epsilon^{-1} \log^2{\epsilon^{-1}})$ time to find the network ID of any message sender in $\Nc(t)$.
    \label{lemma:findid}
\end{theorem}

(Proof in Appendix \ref{sec:app-proofs-direct-upper})
The proof shows that a variant of Peri achieves the upper bound. 
In particular,  to optimize triangular targeted latency, for a target pair of nodes $(s,t)$, the agent must connect to both $s$ and $t$, so $|U|=2$ in Theorem \ref{lemma:findid}. 
Additionally, the following proposition shows that regardless of the agent's algorithm, we require time at least logarithmic in $1/\epsilon$.



%

\begin{proposition}[Lower Bound]
    \label{prop:lb}
    Consider a \name{} network $\Nc(t) = (\Vc(t), \Ec(t))$.
    Let $\Ac$ denote an adversary defined in Theorem \ref{lemma:findid}. 
    For any $\epsilon \in (0, 1)$, to achieve a probability at least $1 - \epsilon$ that the agent is connected to its target,
    $\Ac$ must spend at least $\Omega(\log \epsilon^{-1})$ time, regardless of algorithm.
\end{proposition}
(Proof in Appendix \ref{sec:app-proofs-direct-lower})
Proposition \ref{prop:lb} suggests that a node aiming to prevent agents from learning its network ID with probability at least $1-\epsilon$, should cycle its logical ID on a timescale of order $\log \epsilon^{-1}$.
Together, Theorem \ref{lemma:findid} and Proposition \ref{prop:lb} show how and when an agent can find a message source in a P2P network.
They are an important missing piece in the solution to optimization problems \eqref{eqn:dtl_opt} and \eqref{eqn:ttl_opt} for targeted latency. 
Next, we discuss how 
the agent ensures successful and persistent peer connections with targets after finding them. 


\subsubsection{Connecting to Targets}
By assumption, the target must have at least $f > 0$ dynamic peer slots, which are permissionless and thus can be accessed by the agent. 
Since these slots are on a FCFS basis, the agent may request an occupied peer slot at an arbitrarily high frequency, 
such that it immediately gets it when the slot becomes available after the old connection is torn down. 
\footnote{
To evade detection due to the frequency of requests,
 an agent can spawn many nodes and split the requests among them.} 
In this way, an agent can effectively make a dynamic peer slot of any target \emph{no longer dynamic}, and constantly dedicated to itself. 

\subsubsection{Latency Manipulation} 
\label{sec:unfair-manipulation}
Even if an agent peers with a target, its latency is lower-bounded due to the physical distance between the agent and the target.
Agents may be able to relocate their nodes geographically to manipulate triangular latency.
Currently, over 25\% of Ethereum nodes are running on AWS \cite{aws}.
An agent can 
deploy nodes on co-located cloud servers to reduce the \latt{} to the level of microseconds. 
For \latrt{} where the source and the destination are at different geolocations, 
the agent can deploy node $a_1$ near the source $s$ and node $a_2$ near the destination $t$, where $a_1$ connects to $s$ and $a_2$ connects to $t$, 
with $a_1$ and $a_2$ connected via a low-latency link. 
\sigm{This increases the odds of path $s \to a_1 \to a_2 \to t$ being shortest, which increases the chance of success} in front-running messages between $s$ and $t$. 

\section{Conclusion}

Motivated by the increasing importance of latency in blockchain systems, 
we have studied the empirical performance of strategic latency reduction methods as well as their theoretical limits. 
We formally defined the notions of direct and triangular latency, proposed a strategic scheme \peri for reducing such latencies, and demonstrated its effectiveness experimentally on the Ethereum network.
%
Our results suggest it is not possible to ensure \fairness in 
unstructured 
blockchain P2P networks.
An open question is how to co-design consensus protocols that account for potential abuse at the network layer. 


\section*{Acknowledgements}

We wish to thank Lorenz Breidenbach from Chainlink Labs and Peter van Mourik from Chainlayer for their gracious help in the setup for our direct-targeted-latency experiments.

G. Fanti and W. Tang acknowledge the Air Force Office of Scientific Research under award
number FA9550-21-1-0090, the National Science Foundation under grant CIF-1705007, as well as the generous support of Chainlink Labs and IC3. L. Kiffer contributed to this project while under a Facebook Fellowship. 




\bibliographystyle{ACM-Reference-Format}
\bibliography{bibfile}

\appendix
\section{Appendix}
\label{sec:appendix}

\subsection{Table of Notations}
\label{sec:app-notations}

\resub{We list the table of notations in Table \ref{tab:notations}. }

\begin{table}[!htb]
    \centering
    \begin{tabular}{cc}
    \hline
        Notation & Description \\
    \hline
        $\Nc$ & Network $\Nc = (\Vc, \Ec)$ \\
        $\Vc$ & Set of nodes in the network \\
        $\Ec$ & Set of edges in the network \\
        $\Sc$ & Set of source nodes in the network \\
        $\Tc$ & Set of destination nodes in the network \\
        $\Wc$ & The space of logical node IDs (e.g., pubkey of a wallet in Ethereum) \\
        $d_{\Nc}(a, b)$ & Distance between a pair of nodes $a$ and $b$ over network $\Nc$ \\
        $w(a, b)$ & Link distance between $a$ and $b$ (dominated by their physical distance) \\
        $\Sb$ & Distribution of network traffic sources with support $\Vc$ \\
    \hline
        $\Lambda(a)$ & End-to-end traveling time of a message to $a$ from a random source $S \sim \Sb$ \\
        $\Lambda_v(a)$ & End-to-end traveling time of a message to $a$ from a fixed source $v$ \\
        $L(a)$ & Direct global latency of agent node $a$ over network $\Nc$ \\
        $L_v(a)$ & Direct targeted latency of agent node $a$ w.r.t. target $v$ over network $\Nc$ \\
        $L_{s, t}(a)$ & Triangular targeted latency of agent node $a$ w.r.t. target pair $(s, t)$ over network $\Nc$ \\
        $L_Q(a)$ & Triangular global latency of agent node $a$ w.r.t. set of target pairs $Q$ \\
        $A_\Nc(U)$ & Adversarial advantage when the agent chooses set of peers $U \subseteq \Vc$ \\
    \hline
        $\phi(v)$ & Score of peer $v$ of agent $a$ during a Peri period \\
        $k$ & Budget of number of peers available to the agent node \\
    \hline
    \end{tabular}
    \caption{Table of Notations.}
    \label{tab:notations}
\end{table}

\subsection{\sigm{An Extension of Theorem \ref{lemma:findid}}}

We present Lemma \ref{thm:poisson} which shows a necessary condition that a Poisson process always satisfies. It is proved in \sref{apdx:pf:thm:poisson}. Note that the Poisson process is not the only type of counting process that satisfies \eqref{eqn:quasilive}. 
For instance, a process with constant inter-arrival time also satisfies \eqref{eqn:quasilive}. 

\begin{lemma}
    \label{thm:poisson}
    A Poisson process $\{ M_u(t),~ t \ge 0 \}$ with rate $\nu$ satisfies that there exist constants $\gamma, \mu > 0$, such that 
    \begin{equation}
        \P[M_u(t + \Delta) - M_u(t) < \gamma \Delta] < \dfrac{\mu}{\Delta }, \quad \forall t, \Delta > 0.
        \label{eqn:quasilive}
    \end{equation}
\end{lemma}


Then, we relax the notion of \name{} network to that of \emph{quasi-\name{}} network by relaxing the Poisson process assumption. 

\begin{definition}
A $(\lambda,\nu)$-\emph{quasi-\name{}} network has the following properties:

\begin{enumerate}[label=(\Roman*),leftmargin=*]
    \item
          All {dynamic connections} have a random duration  $\text{Exp}(\lambda)$.
          
    \item For each node $u \in \Sc$, where $\Sc$ denotes a set of target nodes, we let $\{M_u(t),~ t \in [0, \infty)\}$ denote the counting process of messages that are generated and broadcast by $u$. 
    $M_u(t)$ is a counting process, where there exist constants $\gamma, \mu > 0$, such that 
    \[ 
        \P[M_u(t + \Delta) - M_u(t) < \gamma \Delta] < \dfrac{\mu}{\Delta }, \quad \forall t, \Delta > 0.
    \]
      \label{aspt:activity_ext}
\end{enumerate}
\label{def:p2pnet_ext}
\end{definition}

Finally, we extend Thm. \ref{lemma:findid} to Thm. \ref{thm:findid_ext}, relaxing the Poisson process condition to its necessary condition above.
The proof of Thm. \ref{thm:findid_ext} is in 

\begin{theorem}
    Consider a \emph{quasi-\name{}} network $\Nc(t) = (\Vc(t), \Ec(t))$.
    Let $\Ac$ denote an adversarial agent capable of identifying the logical ID of the sender of any message it received from the network.
    Given a set of target nodes $U$,
    with probability at least $1 - \epsilon$ for any $\epsilon \in (0, 1)$,
    it takes the agent $O(|U|^2 \epsilon^{-1} \log^2{\epsilon^{-1}})$ time to find the network ID of any message sender in $\Nc(t)$.
    \label{thm:findid_ext}
\end{theorem}

\subsection{Proofs}
\subsubsection{Proof of Theorem \ref{thm:hardness}}
\label{sec:app-proofs-hardness}
It is known that the set cover problem is NP-complete \cite{karp1972reducibility}. We take an arbitrary instance of the set cover problem: 

\begin{problem*}
    Given a finite set of elements $\Sigma = \{ \sigma_1, \cdots, \sigma_p \}$ and its subsets $\Gamma_1, \cdots, \Gamma_q$. We aim to find the fewest collection of subsets from $\Gamma_{1:q}$, whose union equals $\Sigma$. 
\end{problem*}

We construct a graph $G = (\Vc, \Ec)$ as below. 
Each element $\sigma_i$ in $\Sigma$ corresponds to a unique node of the same name.
Each subset $\Gamma_j$ corresponds to 2 nodes $\gamma^+_j$, $\gamma^-_j$. 
Finally, a fresh node $c$ is added. 
In other words, $\Vc = \{c\} \cup \bigcup_{i=1}^p \{\sigma_i\} \cup \bigcup_{j=1}^q \{ \gamma^+_j, \gamma^-_j \}$.

\begin{figure}[!htb]
    \centering
    \includegraphics[width=.5\linewidth]{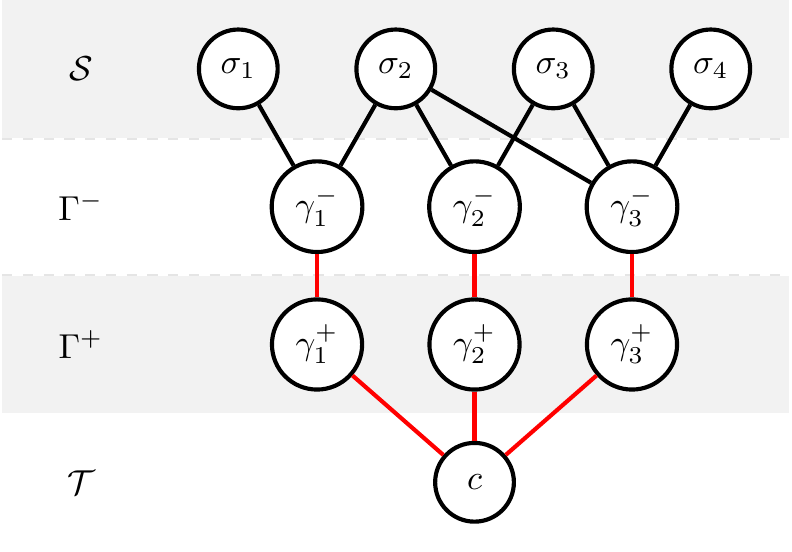}
    \caption{Topology of $G$ given a set cover instance where $\Gamma_1 = \{ \sigma_1, \sigma_2\}, \Gamma_2 = \{ \sigma_2, \sigma_3 \}$ and $\Gamma_3 = \{ \sigma_2, \sigma_3, \sigma_4 \}$.}
    \label{fig:setcover}
\end{figure}

For each pair $(i, j) \in [p] \times [q]$, edge $(\sigma_i, \gamma^-_j) \in \Ec$ if and only if $\sigma_i \in \Gamma_j$. 
Besides these, $\Ec$ contains $(\gamma^-_j, \gamma^+_j)$ and $(\gamma^+_j, c)$ for each $j$. 
\fref{fig:setcover} illustrates an example of topology of $G$.

We take set of sources $\Sc \triangleq \{ \sigma_1, \sigma_2, \cdots, \sigma_p \}$ and set of destinations $\Tc \triangleq \{ c \}$. 
As expected, the original distance $d_G(s, c) = 3$ for each $s \in \Sc$. 
Now we claim that if we can solve Problem \ref{prob:max_adv} in polynomial time for $G, \Sc, \Tc$ where $\forall e \in \Ec ~ \big(  w(e) = 1 \big)$, $\tau = 1.99$ (breaking ties) and $k \in [2, 1+q]$, then we can also solve the original set cover problem in polynomial time. 

When $k \in [2, 1+q]$, we are able to select $k$ spy nodes amongst nodes in $\Vc$. The optimal choice must 
\begin{itemize}[leftmargin=*]
    \item contain $c$: Otherwise, there must exist a spy node in $\Gamma^+$, or the placement will not be effective at all. Replacing this spy node with $c$ will not decrease the advantage. 
    
    \item contain only nodes in $\Gamma^-$ other than $c$: If $\sigma_i$ is chosen, then this spy node serves no other pair than $(\sigma_i, c)$. We pick $j$ where $\sigma_i \in \Gamma_j$ and choose $\gamma_j^-$ instead. This will not decrease the advantage. 
    If $\gamma_j^+$ is chosen, then it benefits the advantage by moving it to $\gamma_j^-$, even when this creates duplicates. 
\end{itemize}

Therefore, to solve the advantage maximization, we are essentially picking nodes in $\Gamma^-$, which corresponds to picking subsets among $\Gamma_1, \cdots, \Gamma_q$. 
If we can solve the advantage maximization in polynomial time for all $k \in [2, q+1]$, we can enumerate all solutions and pick the smallest $k$ where the advantage reaches $p$, the total number of elements in $\Sc$, and also the best advantage we have between $\Sc$ and $\Tc = \{ c \}$. 
This also solves the minimum set cover problem known to be NP-hard, which is a contradiction.
Therefore, Problem \ref{prob:max_adv} is NP-hard, and such polynomial-time algorithm exists only if P $=$ NP. \qeda

\subsubsection{Proof of Proposition \ref{prop:greedy_non_approx}}
\label{apdx:greedy_non_approx}

\begin{figure}
    \centering
    \includegraphics[width=.5\linewidth]{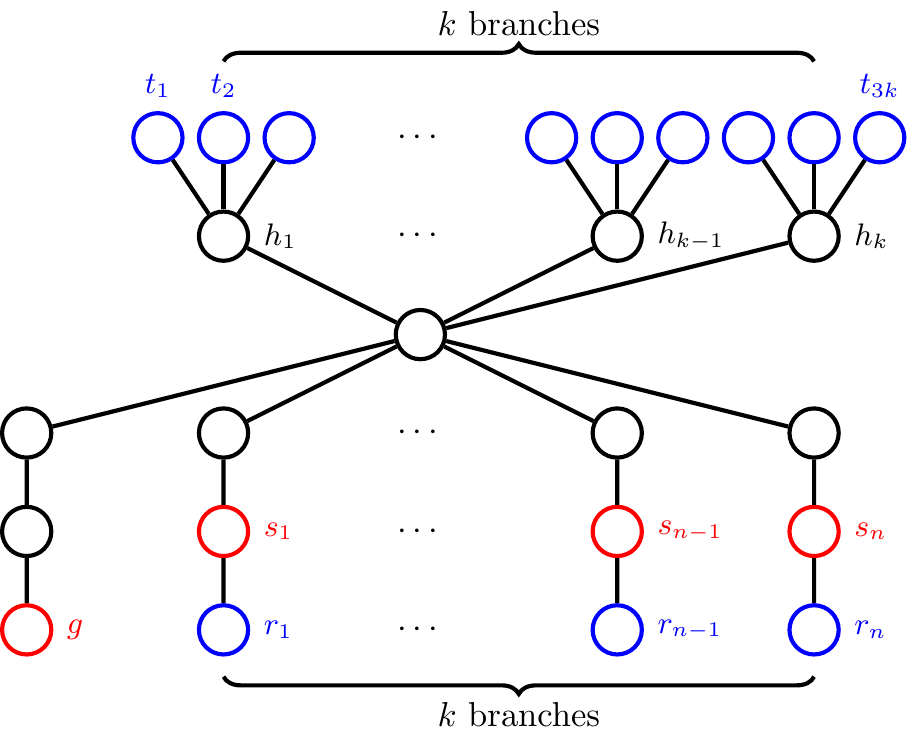}
    \caption{Greedy algorithm cannot maximize advantage with proximity greater than 0. $\tau = 3.99$. Red nodes are sources and blue nodes are destinations. }
    \label{fig:greedy_ce}
\end{figure}

We present the counter-example in \fref{fig:greedy_ce}, which is a tree with $2k+1$ branches.
In this tree, the initial pair of nodes chosen by the greedy algorithm must be $g$ and $h_i$ for some $i \in [k]$,
because a shortcut between them puts 3 pairs at maximum of sources and destinations $\{(g, t_{3i-j}) | j \in \{0, 1, 2\} \}$ under the risk of being front-run.
At each of the following steps, the greedy algorithm will have to continue choosing an additional peer from $\{h_i | i \in [k] \}$, which further increases the advantage by 3.
Hence, with $2\ell$ peers where $\ell < k/2$ is an integer, the total advantage equals $6\ell - 3$.

However, these $2\ell$ peer connections can be put to better uses.
If we choose $\{ s_1, \cdots, s_\ell\} \cup \{ r_1, \cdots, r_\ell\}$ instead, then the shortcut pairs of sources and destinations can be described by
$P = \{ (s_i, r_j) | i \neq j, i \in [\ell], j \in [\ell] \}$, where $|P| = \ell(\ell - 1)$.
Therefore, the ratio of the greedy algorithm solution to the maximum advantage is at most $\frac{6\ell - 3}{\ell(\ell - 1)} \sim \frac{6}{\ell}$,
which can be arbitrarily close to 0 as $\ell, k$ become arbitrarily large.
Equivalently, the greedy algorithm cannot guarantee an $\alpha$-approximately optimal solution for any $\alpha > 0$. \qeda

\subsubsection{Proof of Theorem \ref{thm:findid_ext}}
\label{sec:app-proofs-direct-upper}

We first show the result for a single target, i.e., $|U|=1$.
Let $a$ denote an adversarial node and $x$ denote the message sender (that is, $U=\{x\}$). 
We let the adversary perform the following 3-step strategy iteratively. 

\begin{enumerate}[label=Step \arabic*.]
    \item Wait $\Delta_1$ for the first arrival of a new message $M$ sent by $x$. 
    
    
    
    \item Keep only the peer that contributed to the first arrival, and get $d-1$ random new addresses from the oracle to replace the other peers. The probability of $x$ belonging in these addresses is lower-bounded by $q > 0$. 
    Send peering requests repeatedly until a peer slot becomes available. 
\end{enumerate}

If $x$ is already one of the neighbors at Step 1, it will surely be kept in Step 2 
because $w(x, a)$ is already the shortest path length between $x$ and $a$ on the network, implied by the triangle inequality \eqref{eqn:triangle}. 
In the end, it will remain a peer of the adversary. 
Because the samplings for replacement nodes are independent across $K$ iterations, 
we can geometrically decrease the probability that $x$ is missed in the entire procedure by increasing $K$.
This intuitively explains why $K$ is of order $O(\log \epsilon^{-1})$.
Next, we make this intuition precise. 

Let
\[
    K \triangleq \frac{\log (\epsilon/2)}{\log (1-q)}, \quad \epsilon_1 = \epsilon_2 \triangleq \frac{\epsilon \log (1-q) }{ 4 \log (\epsilon/2)}.
\]

In Step 1, it is desired that the sender sends at least 1 message during the time window of length $\Delta_1$ with probability at least $1 - \epsilon_1$. 



\sigm{By Definition \ref{def:p2pnet_ext}, there exist constants $\mu, \gamma$ such that}
\begin{align*}
    \P[M_x(t + \Delta_1) - M_x(t) \ge \gamma \Delta_1] \ge 1 - \frac{\mu}{\Delta_1}. 
\end{align*}

\sigm{
This assumption essentially states that it is highly probable that during a sufficiently long window of time, a node broadcasts at least $\gamma$ messages every unit of time on average. 
}

Here by taking $\Delta_1 = \max\{ \gamma^{-1}, \mu \epsilon_1^{-1} \}$, we obtain the lower-bound probability that $x$ broadcasts at least 1 message during any time interval of length at least $\Delta_1$. 


In Step 2, we would like to wait $\Delta_2$ for all the $d-1$ new nodes being connected to the adversary. This requires each new node to have at least 1 free slot. 
We consider the worst case where none of the slots are free in the beginning. 
For each target peer $p$, each one of the $F$ slots will become free after a period of time $T_0$, where $T_0 \sim \text{Exp}(\lambda)$. 
Since one slot is already sufficient, the probability the peer $p$ becomes available after $\Delta_2$ is lower-bounded by $1 - e^{-\lambda \Delta_2}$.
This expression is further lower-bounded by $1 - \dfrac{\zeta}{\Delta_2}$ for $\zeta = \dfrac{1}{e\lambda}$. 

Then, considering all the $(d-1)$ peers, we may use the union bound to derive a lower bound of the probability $P_3$ that all the $d-1$ peers are available.
\[ 
    P_3 \ge 1 - \frac{(d-1) \zeta}{\Delta_2}. 
\]

We want $P_3$ to be at least $1 - \epsilon_2$. This can be achieved by letting 
\[
    \Delta_2 = \frac{(d-1) \zeta}{\epsilon_2}.
\]

After taking $K$ iterations of these steps, the probability of finding $x$ equals $1 - (1-q)^K$, while the probability that all executions of Steps 1 \& 3 are successful is at least $1 - K \epsilon_1 - K \epsilon_2$. 
The overall probability equals 
\begin{align*}
    \left[ 1 - (1-q)^K \right] (1 - K \epsilon_1 - K \epsilon_2 ) & \ge 1 - (1-q)^K - K(\epsilon_1 + \epsilon_2) \\ 
    & = 1 - \frac{\epsilon}{2} - \frac{\epsilon}{2} \\
    & = 1 - \epsilon. 
\end{align*}
On the other hand, the total time consumption equals 
\begin{align}
\label{eq:final}
    K \Delta_1 + K \Delta_2 & = O(\log \epsilon^{-1}) \left[ O\left( \epsilon^{-1}\log \epsilon^{-1}\right) + O \left( \epsilon^{-1} \right) \right] \nonumber \\
    & = O\left(\epsilon^{-1} \log^{2}( \epsilon^{-1})\right). 
\end{align}

Next, we show how to extend this result to an arbitrary $U$ via a union bound. 

From \eqref{eq:final}, we know that with probability $1 - \epsilon/|U|$, the adversary is able to find the network ID of an arbitrary node $u \in U$ within time  
\[ 
    O\left((\epsilon/|U|)^{-1} \log^2 (\epsilon/|U|)\right) = O\left(|U| \epsilon^{-1} \log^2 \left(\epsilon^{-1}\right) \right). 
\]

Regardless of how the adversary allocates time to the tasks of finding each $u \in U$, 
the probability of finding all of them is at least $1 - |U| \times \epsilon / |U| = 1 - \epsilon$ by the union bound. 
As for the time consumption, we consider the worst case where the agent has to run the algorithms for finding each $u \in U$ sequentially.
In this case, the total time consumption is the time consumption above of each single task multiplied by number of tasks $|U|$, which equals 
\[
    O\left(|U|^2 \epsilon^{-1} \log^2 \left(\epsilon^{-1}\right) \right). \qeda
\]

\subsubsection{Proof of Proposition \ref{prop:lb}}
\label{sec:app-proofs-direct-lower}
 First of all, we assume $N$, the number of nodes, to be upper-bounded so that the probability $q$ that the oracle returns a target after a single draw satisfies $0 < q_1 \le q \le q_2 < 1$ for some constants $q_1, q_2$. 
In order to connect to the target, it is necessary for the adversary to draw it using the oracle. 
Temporarily, we ignore the other necessary steps (such as judging if an existing peer is the target) and consider only drawing nodes. 
Then, the probability $P$ that the target is drawn within $K$ steps satisfies
\[ 
1 - (1-q_1)^K \le P \le 1 - (1-q_2)^K. 
\]

To let $P = 1 - \epsilon$, we should equivalently have $K = \Theta(\log \epsilon^{-1})$ because
\[ 
\frac{\log \epsilon}{\log (1-q_2)} \le K \le \frac{\log \epsilon}{\log (1-q_1)}. 
\]

Considering that each drawing takes $\Theta(1)$ time, it takes at least $\Theta(K) = \Theta(\log \epsilon^{-1})$ time to find the target with the oracle w.p. at least $1-\epsilon$. 
As argued above, drawing the target is a necessary condition for the final peer connection to it.
Hence, $\Omega(\log \epsilon^{-1})$ is a lower bound of time consumption. \qeda

\subsubsection{Proof of Lemma \ref{thm:poisson}}
\label{apdx:pf:thm:poisson}

    

Let $\nu$ denote the arrival rate. For any $\Delta$, we assign $k = \nu \Delta/2$, and obtain
\begin{align*}
    \P[M_u(t + \Delta) - M_u(t) < \frac{\nu \Delta}{2}] 
    & = e^{-\nu \Delta} \sum_{i=0}^{\floor{\nu \Delta / 2}} \frac{(\nu \Delta)^i}{i!} \\
    & \sim e^{-2k} \sum_{i=0}^k \frac{(2k)^i}{i!} \\
    & \lesssim k e^{-2k} \sup_{t \in [0, k]} \frac{(2k)^t}{t!} \\
    & \lesssim k e^{-k}  \tag{*} \label{eqn:sumred} \\
    & \lesssim \frac{1}{k} \sim \frac{1}{\nu\Delta}.
\end{align*}

This already justifies the claim. It remains to prove \eqref{eqn:sumred}. By Stirling's approximation, let 
\[ 
    f(t) = \log \frac{(2k)^t}{t!} \approx t \log(2k) - \frac{\log t}{2} - t \log t + t. 
\]

As a result, 
\[ 
    f'(t) \approx \log(2k) - \frac{1}{2t} - \log t. 
\]

As $\log t + 1/(2t)$ monotonically increases in $(1/2, \infty)$, we may assert that the maximizer $\tau$ satisfies
\[ 
    f'(\tau) = 0 \eqv 2k = \tau e^{\frac{1}{2\tau}}. 
\]

It can be confirmed that $\tau \in [0, k]$. Hence, 
\[
    \sup_{t \in [0, k]} \frac{(2k)^t}{t!} \sim \frac{(\tau e^{\frac{1}{2\tau}})^\tau}{\sqrt{\tau} \tau^\tau e^{-\tau}} \lesssim e^{\tau} \lesssim e^k. \qeda
\]

\subsection{Statistical Significance of Direct Global Latency Measurements}
\label{app:stats}
To establish the statistical significance of the mean difference among our \latg{} distributions measured in Section \ref{sec:dml:global} over the Ethereum mainnet, we first perform the Kruskal-Wallis H test~\cite{kruskal}. The resulting p-value equals 0, so a subsequent Dunn's test is recommended.
%
We perform Dunn's test \cite{dunn} with Bonferroni correction~\cite{bonferroni}, and obtained a 0 p-value between each pair of distributions. 
This test supports statistically significant differences in the reduction of \latg{} effected by the three methods, ordered by the means of their corresponding distributions.

\subsection{Evaluation of \latt{} over Ethereum Mainnet}
\label{sec:exp-targeted-mainnet}
We ran limited experiments on targeted latency by measuring transactions from a target / victim node in Germany operated by 
Chainlink, an organization providing widely used blockchain services 
; the node sends 2 transactions per hour on average.
These transactions originate from an account (Ethereum address) exclusive to the victim node, i.e.,
all transactions are sent only by the victim.
The node kept running on the network during our experiments.

We evaluate each of the four methods from \sref{sec:baselines}, measuring both their ability to reduce targeted latency and their ability to connect to (i.e., infer) the IP address of our target node on the Ethereum main P2P network.
To this end, we establish 4 EC2 instances in the ap-southeast-1 AWS data center in Singapore, which have a 160 ms round-trip time to Germany.
They are located within the same subnet as a public bloXroute relay.
As in \sref{sec:dml:global}, we deploy a full Go-Ethereum node on each host with at most 50 peers.
The proportion of outbound peers remain the same.
We conduct the experiments under the same procedure and take similar anti-biasing measures, except for three key differences.
First, we define the relevant transactions as only those sent by the target's account.
Second, we set the \peri period to 30 minutes to match the frequency of source transactions.
Third, the duration of each experiment is extended from 8 hours to 16 hours to permit a larger number of \peri periods.

\begin{figure}[htpb]
    \includegraphics[width=.6\linewidth]{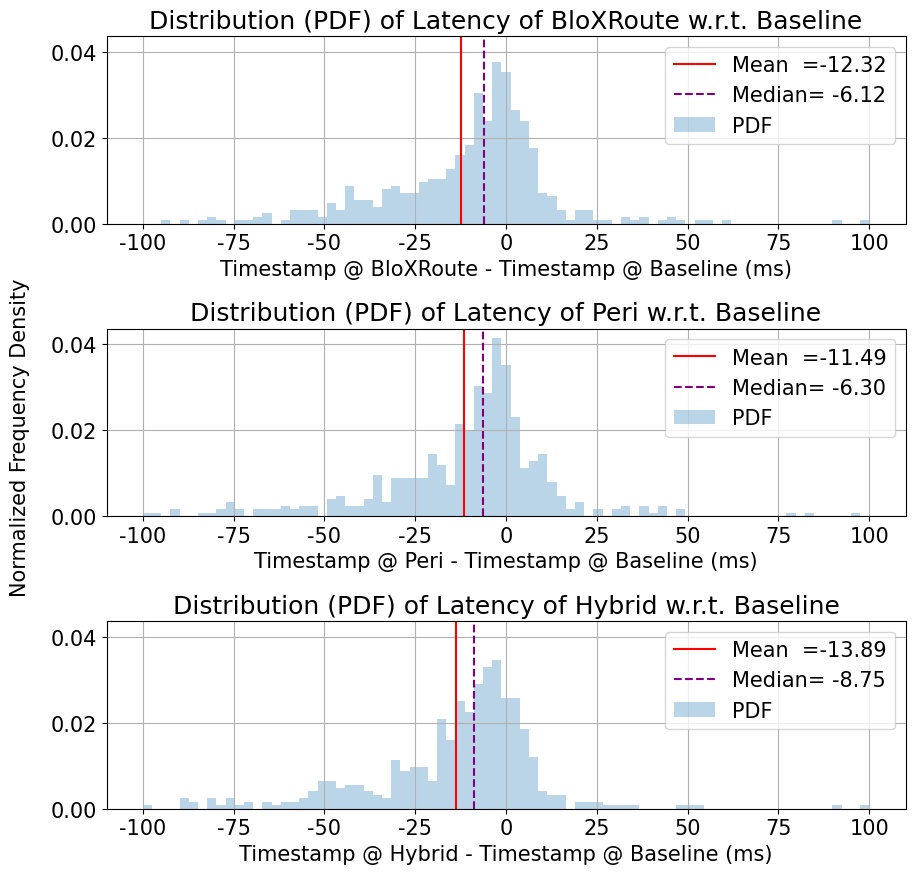}
    \caption{PDFs of distributions of targeted latency. 
    }
    \label{fig:latt_mainnet}
\end{figure}

We ran 23 experiments from March 19, 2022 to April 19, 2022.
The Hybrid and Baseline clients were able to establish connections to the victim node once each.
The distributions of targeted latency of each method are displayed in \fref{fig:latt_mainnet}.
In total, we analyzed the latencies of 654 transactions. 


Over distributions in \fref{fig:latt_mainnet}, we performed Kruskal-Wallis test and obtained p-value~$=0.0235$, 
which allows us to reject the null hypothesis at significance level 5\% and continue with a Dunn's test. 
Then, we performed Dunn's test with Bonferroni correction over these distributions. 
The p-values are listed in Table \ref{tab:pval}. 
At significance level 5\%, we cannot assert a difference between the means of any pair of distributions. 
This indicates that all the methods share a similar ability to reduce \latt{}.
We can further observe this phenomenon from \fref{fig:latt_mainnet}, where they share a similar ability to reduce \latt{} by 14\% to 18\% of the end-to-end delay,
and the hybrid method outperforms the other two methods with a slight advantage.
Unlike reduction of \latg{}, the \peri algorithm is no longer significantly worse than the bloXroute relay network at reducing \latt{},
which makes it a free replacement of bloXroute services for agents with specific targets.
An agent with bloXroute services can also further boost the latency reduction by an additional 13\% by stacking the \peri algorithm and turning hybrid.
In addition, \peri can potentially help an agent identify the IP address of a victim, while bloXroute, which intermediates connections, cannot support such functionality.

\begin{table}[htpb]
    \centering
    \begin{tabular}{cccc}
    \hline
        \textbf{Pair} & (B, P) & (B, H) & (P, H) \\
    \hline
        \textbf{p-value} & 1.0 & 0.0532 & 0.0529 \\
    \hline
    \end{tabular}
    \caption{
        Pairwise p-values of Dunn's test over distributions of targeted latency in \fref{fig:latt_mainnet}, comparing bloXroute (B), \peri (P), and Hybrid (H). 
        \vspace{-4ex}
    }
    \label{tab:pval}
\end{table}


\subsection{\sigm{Evaluation of Relation Between \latg{}{} and Peer Count}}
\label{sec:eval_pc}

\sigm{
We ran additional experiments for comparing \latg{} when we vary the peer count of the adversary. 
We use 2 EC2 (i3.xlarge) instances in the us-east-1 AWS data center, where each instance deploys a full node with a customized Go-Ethereum (geth) client, built on top of the nemata (1.10.25) release.
To correctly run the nodes after the Ethereum merge in September 2022, we ran a lighthouse client\footnote{\url{https://github.com/sigp/lighthouse/releases/tag/v3.3.0}} at default settings in parallel. 
We vary the maximum peer count in the range from 25 to 200. 
For each peer count, we ran 12 experiments in January 2023 with each following the procedure below. 
Each experiment lasted 8 hours. 
We adaptively selected the period of Peri for each peer count to let the Peri node accumulate enough candidates to choose from. 
The experiments were configured by the same workflow as described in \sref{sec:dml:global}. 
For financial reasons, we did not test bloXroute and Hybrid methods with 2 additional AWS nodes.
}

\begin{figure}
    \centering
    \includegraphics[width=\linewidth]{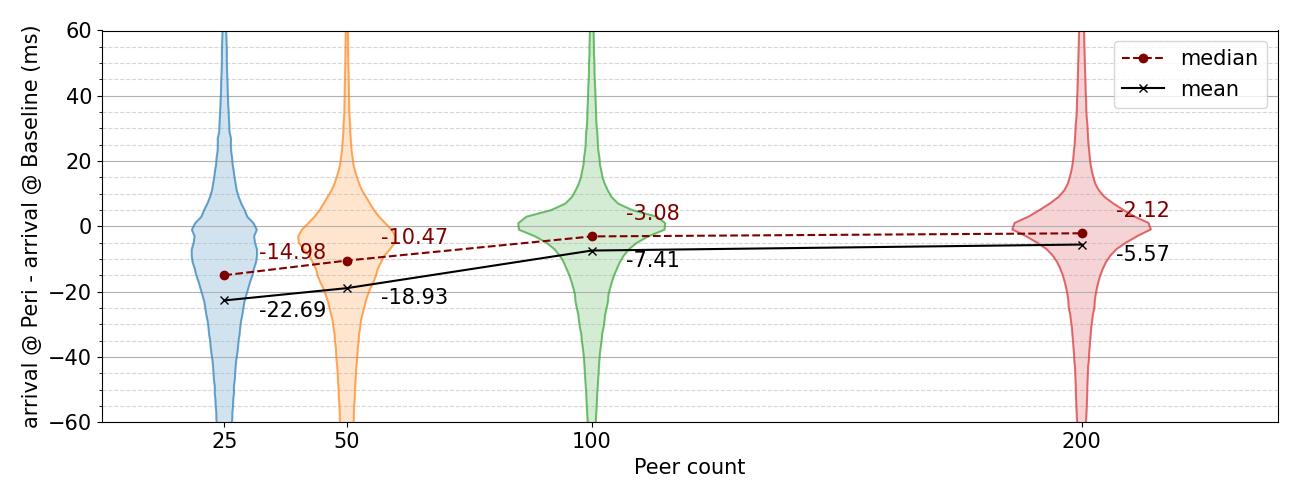}
    \caption{PDFs of global latency with maximum peer counts 25, 50, 100 and 200. The PDFs are normalized by the same factor for legibility. }
    \label{fig:latg_peercount}
\end{figure}

\sigm{
Each node is allowed to warm up for 2.5 hours; after this, we collect all transactions that are received by each of the nodes for 3.5 hours. 
As we did in \sref{sec:dml:global}, we collected the samples of random variable $\Lambda(y) - \Lambda(b)$, which represents the arrival time difference of one transaction $m$ between at the Peri node $y$ and the baseline node $b$. 
The smaller the time difference (i.e., the more negative), the earlier $y$ delivers $m$, and the more effective Peri is.
For each peer count, we gather the latency differences over 12 experiments, and plot their distributions for each node in \fref{fig:latg_peercount}.
Each distribution is estimated over 500,000 transactions.
}

\sigm{
In \fref{fig:latg_peercount}, we observe that Peri maintains an advantage over the baseline, regardless of the peer count.
However, the more peers the node maintains, the less advantage Peri has, and the more concentrated the distribution is around 0. 
It is still notable that Peri has a 5.5ms advantage with a large peer count of 200. 
}



\subsection{\sigm{Performance of \peri under Limited Bandwidth}}
\label{sec:eval_bw}

\sigm{
We ran additional experiments for comparing \latg{} under different networking conditions. 
Our nodes were deployed in the same way as in App. \ref{sec:eval_pc}. 
We set the peer count to 50, the default option of Go-Ethereum.
We limited the bandwidth of both nodes by setting the bandwidth limit of both inbound and outbound traffic.
In all experiments where the bandwidths were unlimited, the machines had ``up to 10 Gbps'' bandwidth. 
\footnote{The speed tests typically report 1 to 2 Gbps.}
For each bandwidth, we ran 10 experiments in January 2023 with each lasting 4.8 hours. 
The experiments were configured by the same workflow as described in \sref{sec:dml:global}. 
For financial reasons, we did not test bloXroute and Hybrid methods with 2 additional AWS nodes.
}

\begin{figure}
    \centering
    \includegraphics[width=\linewidth]{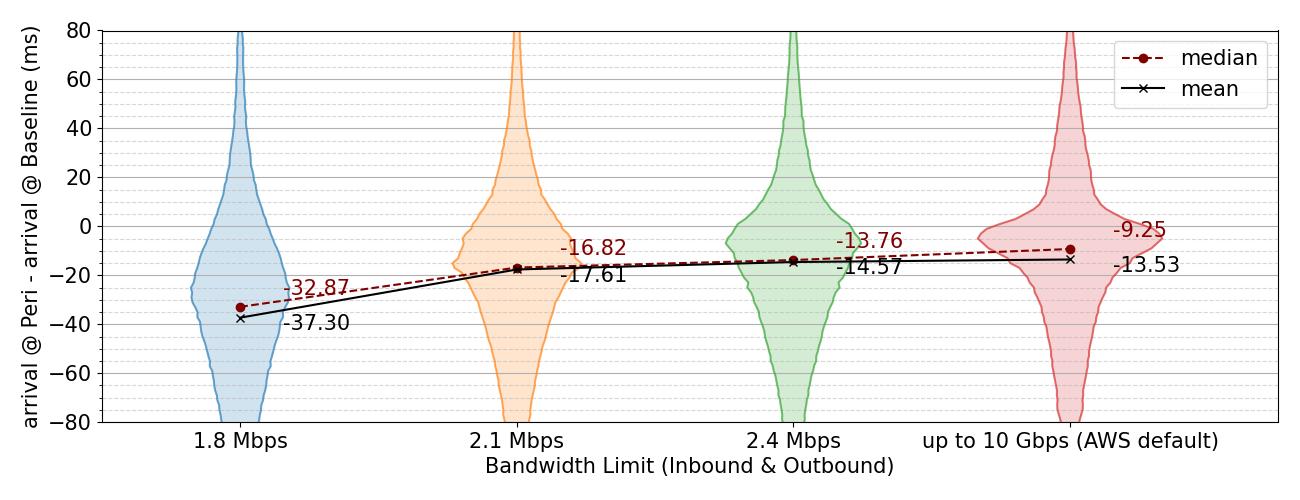}
    \caption{
        PDFs of distributions of global latency under limited bandwidth.     
    }
    \label{fig:latg_bw}
\end{figure}

\sigm{
Each node is allowed to warm up for 2.5 hours; after this, we collect all transactions that are received by each of the nodes for 2.3 hours. 
The data is collected in the same way as we did in \sref{sec:dml:global} and App. \ref{sec:eval_pc}. 
For each bandwidth, we gather the latency differences over 10 experiments, and plot their distributions for each node in \fref{fig:latg_bw}.
Each distribution is estimated over 350,000 transactions.
}

\sigm{
In \fref{fig:latg_bw}, we observe that the lower bandwidth an agent has access to, the higher advantage Peri gains over the baseline.
}

\bigskip

\subsection{\sigm{Full Result of Simulations of Advantage Maximization Algorithms on Hub-Enriched Topologies}}
\label{sec:sim_cent_all}

\sigm{
    We present the extended results of \sref{fig:sim_cent_top} and a full description of our simulation setups, which we lack space to show in the main text. 
    Cryptocurrency P2P topologies are notoriously difficult and expensive to measure \cite{delgado2019txprobe}, and to our knowledge, there are no public datasets of recent cryptocurrency P2P topologies \cite{coinscope,delgado2018cryptocurrency,deshpande2018btcmap,eisenbarth2021open}. 
    As a compromise, we took 2 snapshots of cryptocurrency topologies in real life -- the Bitcoin P2P network topology on 9/4/2015 \cite{miller2015discovering} and the Lightning network topology in 8/23/2022 \cite{lngossip}. The Bitcoin data is not publicly available, and was obtained by contacting the authors of \cite{miller2015discovering}.
    The LN snapshot was directly collected \cite{lngossip}.
}
\cam{
    Although the Lightning network is a layer-2 network where latency wars do not typically take place, it is still a P2P network and its topology may have some similarities to layer-1 P2P networks. Therefore, we add simulation results on the Lightning network topology for reference. 
}
\sigm{
    The major component of the Bitcoin P2P network consists of 4,654 nodes and 18,467 edges, and that of the Lightning network consists of 36,553 nodes and 296,589 edges. 
    For a fair comparison against the random topologies, we generate 25 sub-topologies of each topology by the snowball-sampling algorithm.  
    All subgraphs consist of 300 nodes, but their average node degree differ. 
    The average degrees of Bitcoin subgraphs range from 2.3 to 4.9, while those of Lightning subgraphs range from 3 to 35. 
    \fref{sec:sim_cent_all} shows the performance in advantage metric of 3 different methods over 4 random synthetic topologies and 2 topologies in practice. 
}

\begin{figure}[!htb]
    \centering
    \includegraphics[width=\linewidth]{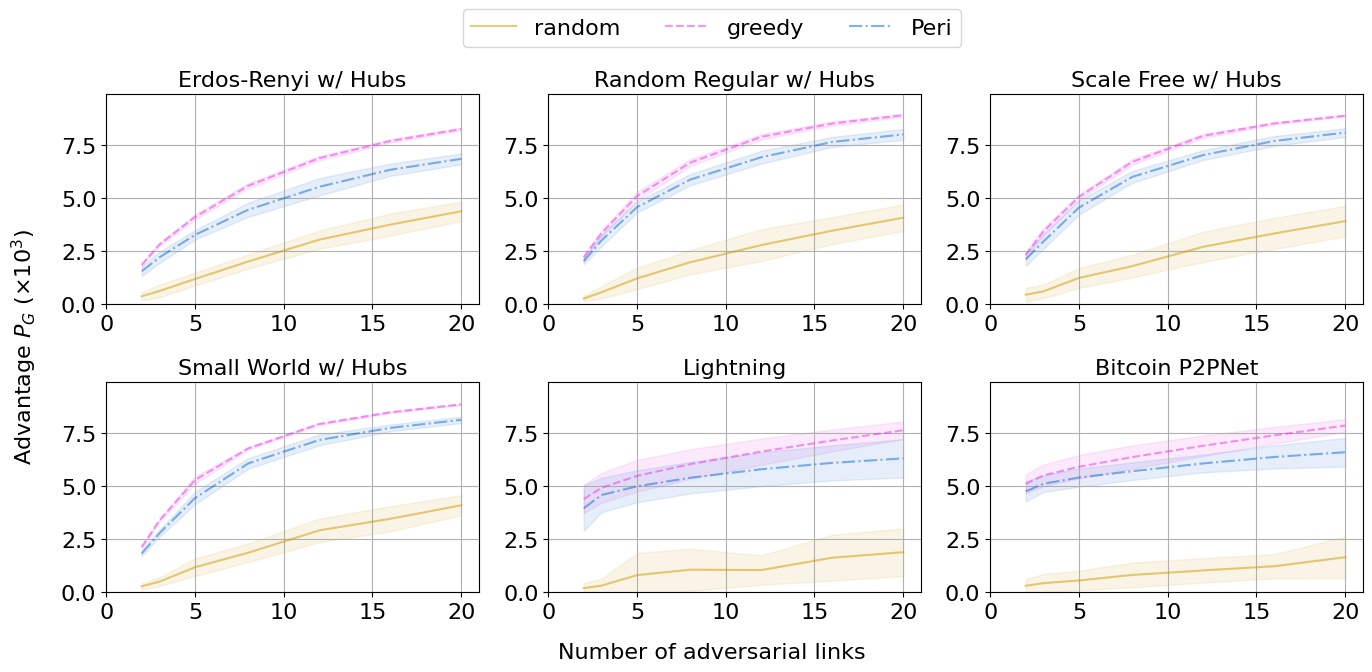}
    \caption{Full version of \fref{fig:sim_cent_top}. The mean of the curves are shown by solid lines, and the standard deviations are shown by the transparent color zones centered at the mean. }
    \label{fig:sim_cent_top_all}
\end{figure}

\subsection{Simulations of Advantage Maximization Algorithms on Original Topologies}
\label{sec:sim_orig}

\begin{figure}
    \centering
    \includegraphics[width=\linewidth]{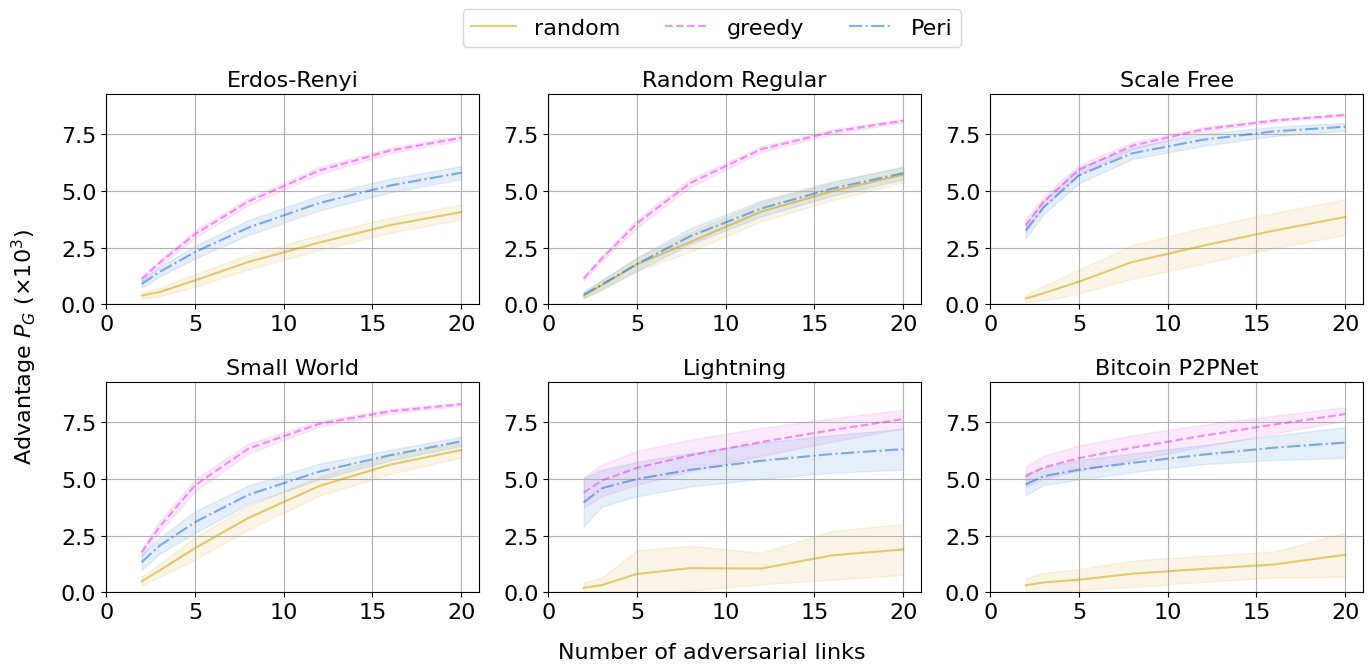}
    \caption{Advantage-Peer-count curves on original random graph models. The mean of the curves are shown by solid lines, and the standard deviations are shown by the transparent color zones centered at the mean. \vspace{-2ex}}
    \label{fig:sim_orig_top}
\end{figure}

We synthesize the network models as in \sref{sec:triangular-approximations} without centralizing them by introducing hub nodes.
We reuse other setups in the original simulation, and plot the advantage-peer-count curves in \fref{fig:sim_orig_top}.

For all the graph models, both \peri and Greedy achieve a higher advantage $A_{\Nc}$ than the random baseline,
with Greedy outperforming \peri by varying amounts.
For the most decentralized models, such as the random regular and small world, the advantage of \peri is much closer to that of random baseline than the greedy algorithm.
On the other hand, for models with a few high degree nodes (i.e., hubs), \peri inserts shortcut peering connections almost as well as the greedy algorithm, in spite of its limited knowledge of the graph.
Therefore, the performance of \peri is likely to be stronger on networks with many hubs.
This is consistent with our conclusion in \sref{sec:triangular-approximations}.

\end{document}